\documentclass[fleqn,usenatbib]{mnras}
\usepackage{newtxtext,newtxmath}
\usepackage{soul}
\usepackage[T1]{fontenc}
\DeclareRobustCommand{\VAN}[3]{#2}
\let\VANthebibliography\thebibliography
\def\thebibliography{\DeclareRobustCommand{\VAN}[3]{##3}\VANthebibliography}

\usepackage{graphicx}	
\usepackage{amsmath}	
\usepackage{multirow}
\usepackage{hyperref}

\title[Spatial Evolution of Stellar Clusters in NGC\,628]{The Spatial Evolution of Star Clusters in NGC\,628 with JWST}

\author[Anne S. M. Buckner et al.]{
Anne S. M. Buckner,$^{1}$\thanks{E-mail: bucknera@cardiff.ac.uk}
Ana Duarte-Cabral,$^{1}$
Angela Adamo,$^{2}$
Sean Linden,$^{3}$
Michele Cignoni,$^{4,5,6}$
\newauthor
Varun Bajaj,$^{7}$
Arjan Bik,$^{2}$
Giacomo Bortolini,$^{2}$
Daniela Calzetti,$^{8}$
Matteo Correnti,$^{9,5}$
Bruce G. Elmegreen,$^{10}$
\newauthor
Debra M. Elmegreen,$^{11}$
Helena Faustino Vieira,$^{2}$
John S. Gallagher III,$^{12}$
Kathryn Grasha,$^{13,14}$
\newauthor
Benjamin Gregg,$^{15}$
Rob A. Gutermuth,$^{8}$
Kelsey Johnson,$^{16}$
Mark Krumholz,$^{13}$
Drew Lapeer,$^{8}$
\newauthor
Matteo M. Messa,$^{5}$
Göran Östlin,$^{2}$
Alex Pedrini,$^{2}$
Jenna E. Ryon,$^{7}$
Linda J. Smith,$^{17}$
Monica Tosi$^{5}$
\\
\\
$^{1}$Cardiff Hub for Astrophysics Research and Technology (CHART), School of Physics \& Astronomy, Cardiff University, The Parade, CF24 3AA Cardiff, UK\\
$^{2}$Department of Astronomy, Oskar Klein Centre, Stockholm University, AlbaNova University Center, SE-106 91 Stockholm, Sweden\\
$^{3}$Steward Observatory, University of Arizona, 933 N Cherry Avenue, Tucson, AZ 85721, USA\\
$^{4}$Dipartimento di Fisica, Università di Pisa, Largo Bruno Pontecorvo 3, 56127 Pisa, Italy\\
$^{5}$INAF – OAS, Osservatorio di Astrofisica e Scienza dello Spazio di Bologna, via Gobetti 93/3, I-40129 Bologna, Italy\\
$^{6}$INFN, Largo B. Pontecorvo 3, 56127 Pisa, Italy\\
$^{7}$Space Telescope Science Institute, 3700 San Martin Drive, Baltimore, MD 21218, USA\\
$^{8}$Department of Astronomy, University of Massachusetts, 710 North Pleasant Street, Amherst, MA 01003, USA\\
$^{9}$ASI-Space Science Data Center, Via del Politecnico, I-00133, Rome, Italy\\
$^{10}$Katonah, NY 10536, USA\\
$^{11}$Vassar College, Department of Physics \& Astronomy, Poughkeepsie, NY 12604, USA\\
$^{12}$Department of Astronomy, University of Wisconsin, 475 North Charter St., Madison, WI 53706, USA\\
$^{13}$Research School of Astronomy and Astrophysics, Australian National University, Canberra, ACT 2611, Australia\\
$^{14}$ARC Centre of Excellence for All Sky Astrophysics in 3 Dimensions (ASTRO 3D), Australia\\
$^{15}$University of Massachusetts Amherst, Amherst, MA 01003, USA\\
$^{16}$Department of Astronomy, University of Virginia, Charlottesville, VA, USA\\
$^{17}$Space Telescope Science Institute, 3700 San Martin Drive, Baltimore, MD 21218, USA
}

\date{Accepted XXX. Received YYY; in original form ZZZ}

\pubyear{\the\year{}}

\begin{document}
\label{firstpage}
\pagerange{\pageref{firstpage}--\pageref{lastpage}}
\maketitle

\begin{abstract}
We examine the spatial distribution of star clusters in NGC628 using the statistical tool INDICATE to quantify clustering tendencies. Our sample, based on HST and JWST observations, is the most complete to date, spanning ages from 1\,Myr to $>$100 Myr. We find cluster spatial behavior varies with galactic position, age, and mass. Most emerging young clusters are tightly spatially associated with each other, while fully emerged clusters are in $\sim$1.5 times looser spatial associations, irrespective of age. Young Massive Clusters (YMCs\,$\ge10^{4}$\,M$_{\odot}$) tend to associate with lower-mass clusters but not strongly with other YMCs, implying that intense star formation regions produce a few YMCs alongside many lower-mass clusters rather than multiple YMCs together. Young concentrated clusters show a wide radial distribution in the galactic disk, which narrows with age; with concentrated clusters $>$\,100\,Myr mostly residing between 2–6 kpc. This pattern may reflect either faster dispersal of isolated tight cluster spatial `structure' in a lower gas density outer disk or gradual inside-out growth, with the formation of this structure shifting outward over time. We also detect distinct spatial behaviors for clusters within 2\,kpc, linked to the inner Lindblad resonance ($\le$1\,kpc), nuclear ring ($\sim$0.5–1\,kpc), and the start of spiral arms ($\sim$1.25–2\,kpc), suggesting these regions exhibit strong radial motions that could hinder clusters from forming and remaining in tight concentrations. Our results highlight how spatially-resolved studies of clusters can reveal the influence of galactic dynamics on star formation and cluster evolution. 

\end{abstract}



\clearpage
\section{Introduction}


Star clusters are the fundamental building blocks of spiral galaxies as most stars form in clustered environments. Unfortunately, the mechanisms of cluster formation, effect of local and galactic environmental conditions have on this, as well as their longer-term evolution, remain unclear despite considerable efforts by both the theoretical and observational communities. To better constrain the role of galactic environment requires a multi-pronged analysis of cluster populations, comparing variation in their properties (age, mass, stellar composition) as a function of galactic environment and conditions. Young Massive Clusters (YMCs) – characterised as having ages $<100$\,Myr and masses $\ge10^{4}$M$_{\odot}$ - are of particular interest as they provide important clues about the phyiscs underpinning massive cluster and star formation.  Thus, when examined as part of larger cluster population studies, the influence of the galactic environment on both the early and longer-term evolution of massive clusters can be explored. The advent of high-resolution infrared data from the James Webb Space Telescope (JWST; \citealt{ 2006SSRv..123..485G}) provides the opportunity to greatly advance our understanding of these processes, as many of the youngest clusters – still associated with their natal clouds – are observable for the first time. 

Here, we characterise the spatial properties of the star cluster population of spiral galaxy NGC\,628. Also known as Messier 74, it is a multiple arm galaxy of type SA(s)c with two main spiral arms and is the principal galaxy of the M74 galaxy group. NGC\,628 has an estimated total stellar mass of $2.2 \times 10^{10} M_{\odot}$ \citep{2021ApJS..257...43L}, is nearby (9.84\,Mpc; \citealt{2009AJ....138..323T}), bright ($V=9.46$\,mag; \citealt{2014MNRAS.445..881C}), face-on ($i=8.9^{\circ}\pm 12.2^{\circ}$; \citealt{2021ApJS..257...43L}), and is actively forming stars with an estimated star formation rate (SFR) of $1.74\,M_{\odot} yr^{-1}$ (\citealt{2021ApJS..257...43L}). NGC\,628 has several multi-wavelength datasets available, including recent James Webb Space Telescope (JWST; \citealt{Gardner_2023}) observations, making it, unsurprisingly, one of the best studied nearby spiral galaxies. It has had studies devoted to its star formation history and efficiency as a function of environment (\citealt{2016ApJ...827..103K}, \citealt{2018ApJ...863L..21K}, \citealt{2022MNRAS.517.3763L}, \citealt{2023ApJ...944L..20K}, \citealt{2024PASJ..tmp...76D}); the distribution and properties of giant molecular clouds (\citealt{2025MNRAS.tmp..403V}, \citealt{2020A&A...634A.121H}, \citealt{2023atyp.confE..46D}); regularly-spaced 8$\mu$m cores along the spiral arms, HII regions,  Polycyclic Aromatic Hydrocarbons and dust (\citealt{2019ApJS..245...14E}, \citealt{2018MNRAS.477.4152R},  \citealt{2023ApJ...944L..23D}, \citealt{2023ApJ...944L..16E}, \citealt{2023ApJ...944L..13T}, \citealt{ 2024ApJ...971...32P}, \citealt{2024ApJ...971..115G}, \citealt{2024ApJ...971..118C}, \citealt{2024arXiv240202659M}); feedback-driven bubbles (\citealt{ 2023ApJ...944L..22B}, \citealt{2023ApJ...944L..24W}); magnetic structure (\citealt{2017A&A...600A...6M}, \citealt{2022A&A...665A..64W}, \citealt{2024ApJ...967...18Z}); and the broader structure/evolution of the galaxy (\citealt{2018MNRAS.476.1909A}, \citealt{2018MNRAS.476.3591M}, 
\citealt{2021MNRAS.506...84I}, \citealt{2022MNRAS.516.2171U}). 

Numerous studies have focused on the demographics of NGC\,628’s star cluster population, extracting key characteristics using a wide-variety of statistically-based approaches. For example, both box-counting and surface brightness threshold methodologies have shown that the cumulative size distribution function of young stellar regions is a power law with a slope of approximately -1.5 (\citealt{2006ApJ...644..879E}, \citealt{2014MNRAS.442.3711G}), while Bayesian modelling has revealed the absence of mass-dependent disruption in the galaxy \citep{2024MNRAS.532.4583T}. The use of cumulative distribution functions found that there is no age gradient across the spiral arms \citep{2018MNRAS.478.3590S} but clusters do display age-dependent spatial behaviour, such that the youngest strongly clustered together but the older population is non-clustered, as identified using the two-point correlation function (\citealt{2015ApJ...815...93G}, \citealt{2017ApJ...840..113G}, \citealt{2018MNRAS.478.3590S}). 
Although these statistics are distinct, they are also similar in that they look at the ‘big picture’ trends across the population i.e. a single parameter is derived for a group of clusters, rather than individual clusters. Advantageously this allows typical galaxy-scale behavioural trends to be described (as a function of average age, mass etc). However, by design they do not describe the behaviours of individual clusters within each group, so cannot identify local-scale variances in behaviour, nor quantify how various physical factors have impacted these. In this paper we will use the novel statistical clustering tool “INDICATE” (INdex to Define Inherent Clustering And TEndencies; \citealt{2019A&A...622A.184B}) to describe the spatial behaviour of each cluster in NGC\,628’s population individually. We will quantify, for the first time, how small differences in cluster age, mass, and galactic position influence spatial behaviour of stellar clusters in NGC\,628. Our study includes a significant number of recently discovered clusters still embedded within their natal clouds, which were found with high-resolution Near InfraRed (NIR) JWST images of NGC\,628 obtained by the Feedback in Emerging extragAlactic Star ClusTers (FEAST, GO 1783, PI A.\,Adamo) collaboration.

This paper is structured as follows. In Section\,\ref{sect_data} we describe our cluster catalogue, and in Section\,\ref{sect_method} we detail the methodology used to study their spatial properties. The results of our analysis for clusters' spatial behaviour, including trends with galactic position, age, and mass, are presented in Section\,\ref{sect_results}, and our conclusions given in Section\,\ref{sect_conclude}.

\begin{figure*}
\includegraphics[width=\textwidth]{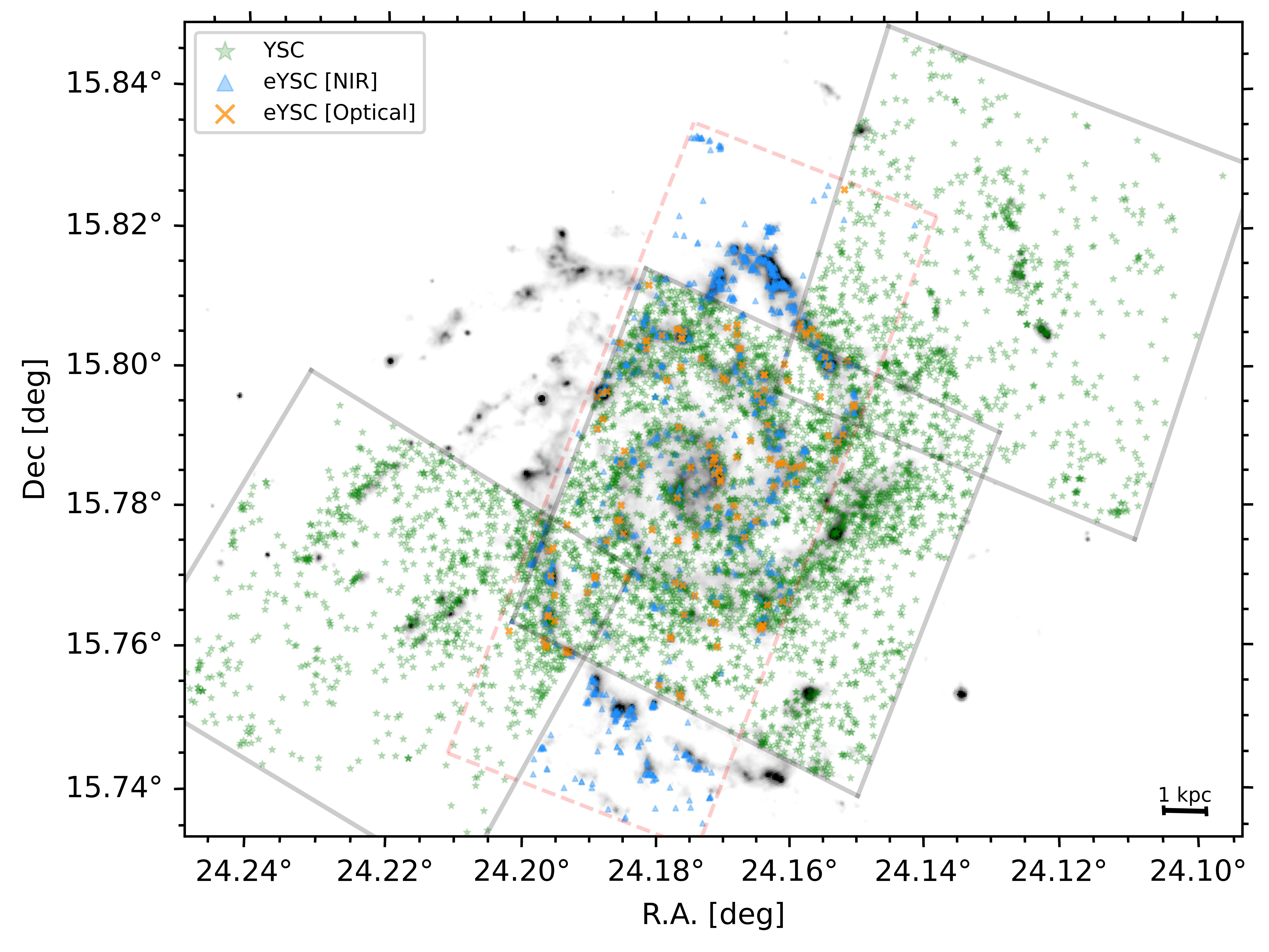}
\caption{Final cluster sample as described in Sect.\,\ref{sect_final_sample}, overlaid on the Spitzer Infrared Nearby Galaxies Survey $8\mu$m image of NGC\,628. Coloured markers denote which catalogue the cluster originates from, as specified by the legend. The grey and red-dashed panels denote the position of the HST and JWST images respectively.  \label{fig_cat_radec}}
\end{figure*}


\section{Data}\label{sect_data}

In this work we analyse the clusters of NGC\,628, using a combined sample consisting of candidates identified using both HST and JWST observations. Below we describe how our sample was constructed and detail our selection criteria. 





\subsection{Initial Sample Selection}\label{sect_cat_selection}
To ensure a complete (as possible) stellar cluster population for our analysis, we created our sample by combining the star cluster catalogues produced within the FEAST collaboration. The first catalogue consists of young clusters in the emergence phase (emerging young stellar clusters, eYSCs) which have been extracted in JWST/NIRCam imaging. The second catalogue contains stellar clusters extracted in optical HST imaging using a well established approach that has led to the production of many star clusters catalogues in local Universe galaxies over the years (see e.g \citealt{2019ARA&A..57..227K} for a review). Hereafter, we will refer to the sample of clusters extracted in this way as optical YSCs for simplicity, although we note that these optical clusters are not necessarily young. The two samples combined cover the entire cluster population from the earliest phases to clusters $>100$\,Myr old.

\subsubsection{eYSC Catalogue}\label{sect_cigale}

The eYSC catalogue presented in Adamo et al. (subm.) consists of 1576 emerging cluster candidates which are still associated with their HII regions and photo-dissociation regions. They were extracted from JWST Cycle 1 observations under the FEAST program (GO1783), with additional archival data from program GO2107 (PI J.\,Lee) included. We opted to utilise this catalogue for our analysis as it represents the most complete dedicated emergence stage cluster catalogue for NGC\,628 to date. Full descriptions of observations, cluster identification, classification, and fitting procedures will be made available in Adamo et  al. (in prep), Linden et al. (in prep) and \citet{2025ApJ...992...96P}. In brief, the GO1783 observations were carried out in the F115W, F150W, F187N, F200W, F277W, F335M, F405N, F444W, and GO2107 observations in F300M and F360M filters. Data was processed using the calibration files 1.11.4, and the final mosaics have been aligned to Gaia and have a final pixel scale of 0.04"/px. eYSCs were identified as candidates which have compact morphologies in Pa$\alpha$, and if present, in Br$\alpha$. Further differentiation was made between eYSCs of two evolutionary classes, Type 1 (eYSCI) and Type 2 (eYSCII), based on whether they have compact morphologies in 3.3 $\mu$m Polycyclic Aromatic Hydrocarbons (PAH) or not, respectively (\citealt{2024ApJ...971...32P}, \citealt{2024ApJ...971..115G}). This classification resulted in 835 and 741 entries respectively. Physical properties were obtained for the eYSCs using SED fitting with CIGALE \citep{2019A&A...622A.103B}. CIGALE operates by generating a grid of models based on the user's input parameters, comparing a cluster's photometry with each model of the grid, and calculating the $\chi^{2}$ value to determine the goodness-of-fit. The $\chi^{2}$ value is then converted into a likelihood via $\exp(-\chi^{2}/2)$, which is then used as a Bayesian prior estimate for the physical properties of each source, where $1\sigma$ uncertainties are then determined by the likelihood-weighted standard deviation of all the models. Once CIGALE fits a cluster's spectral energy distribution (SED) based on the age and reddening, the mass is appropriately scaled based on the cluster's luminosity assuming a fully-sampled IMF. A Milky Way-like extinction curve from \cite{1989ApJ...345..245C} was assumed, with an extinction factor of $R_{V} = 3.1$, solar metallicity, internal reddening in the range from $E(B{-}V) = 0$ to $5$ mags in 0.01~mag steps, ages from 1~Myr to 10~Myr with $\Delta T=$~1~Myr, and total covering fractions which range from 0.01-1. The adopted upper age limit of \(10\,\mathrm{Myr}\) follows from the FEAST selection, which identifies embedded candidates via strong hydrogen recombination emission (Pa\,\(\alpha\)/Br\,\(\alpha\)); the equivalent widths of the near–IR hydrogen lines (e.g. Pa\,\(\beta\), Br\,\(\gamma\)) decline to negligible levels by \(\sim 10\,\mathrm{Myr}\) in Starburst99 models \citep[e.g.][]{Leitherer1999SB99,Leitherer2014SB99}, so older clusters would not be selected by this tracer.

For our sample, we removed sources detected in less than 4 bands, and/or poor SED fit ($\chi^{2}\,\ge\,20, \le\,200\,M_{\odot}$). This selection leaves 1216 clusters remaining, of which 653/563 are eYSCI/II, respectively.

\subsubsection{Optical YSC Catalogue}\label{sect_ygg}

The FEAST optical YSC catalogue (Adamo et al (subm.)) consists of 16645 cluster candidates which were extracted from archival HST data including the Legacy ExtraGalactic UV Survey (LEGUS; \citealt{2015AJ....149...51C}) Hubble treasury program, GO9796, and GO13773. We used this catalogue for our analysis, as it is the largest extraction of the HST F555W FoV clusters to date, surpassing the previously published reference catalogue of LEGUS \citep{2017ApJ...841..131A}. A comparison of this and the LEGUS catalogue is extensively discussed in Adamo et al (subm.). The FEAST catalogue shows clear improvements over the LEGUS catalogue due to a different adopted aperture correction method and the inclusion of narrow-band filters, capturing H$\alpha$ and Pa$\alpha$, and NIR broadband capturing the stellar continuum. 

We refer the reader to the original catalogue paper for full details of the extraction, classification and fitting procedures. In brief, after the initial extraction was performed, final mosaics in WFC3 F275W, F336W, F657N, and ACS/WFC F435W, F555W, F606W, F658N, and F814W were drizzled to a common pixel scale of 0.04"/px and registered to the Gaia reference frame. Properties for the candidates were derived by fitting their SED from 0.3 to 2 $\mu$m (HST F275W, F336W, F657N, F435W, F555W, F606W, F658N, and F814W and NIRCam F115W, F187N, F200W) with Yggdrasil spectral evolutionary models \citep{2011ApJ...740...13Z}, with a metallicity of $Z=0.02$ and a \citet{2001MNRAS.322..231K} initial mass function. Nebular emission  was included assuming a covering factor of 50\% and $n_e = 100$ cm$^{-3}$. 

For our analysis, we removed sources detected in less than 4 bands, flagged as ‘star or spurious object’, and/or poor SED fit ($\chi^{2}\,\ge\,65, \le\,200M_{\odot}$), resulting in a total of 5806 optical YSCs remaining. 




\subsection{Final Sample}\label{sect_final_sample}

We combine the selections of optical YSCs and eYSCs to create our final sample, made necessary as (i) high extinctions associated with eYSCs mean a significant portion are expected to be below the detection threshold of HST and thus absent from the optical YSC catalogue; (ii) despite there being significant overlap between the IR and optical coverage, due to the JWST mosaic orientation eYSCs are detected in some regions where there is no HST coverage (Fig.\,\ref{fig_cat_radec}). To remove duplicate entries from the two catalogues, a crossmatch is performed with the criteria that any eYSCs within a 4 pixel tolerance ($0.16"$) of an optical YSC (i.e. the aperture radius used to extract photometry) is considered a match. In the cases where more than 1 match is found for a single source, only the closest pairing is considered. A total of 136 matches were found, with four instances of an eYSC being paired with two YSCs, leaving the final combined sample of 6890 clusters. For the cross-matched YSCs we use the properties (age, mass) from the eYSC catalogue. The age and mass ranges for the sample are shown in Fig.\,\ref{fig_cat_agemass} and the top panels of Fig.\,\ref{fig_cat_agemass_hists}. Finally, each cluster is assigned (i) an environmental flag denoting their spatial position in NGC\,628's centre, spiral arms or interarm regions, using the environmental mask of NGC\,628 from \citet{2021A&A...656A.133Q}, as shown in Fig.\,\ref{fig_RaDec_I5} and (ii) an origin flag denoting which catalogue(s) the cluster originates from, namely:

\begin{itemize}
    \item \textit{YSC} - Optical YSC Catalogue
    \item \textit{eYSC [NIR]} - eYSC Catalogue
    \item \textit{eYSC [optical]} -  Optical YSC and eYSC Catalogue.
\end{itemize}

A comparison between the physical properties obtained for clusters using CIGALE and Yggdrasil SED fitting has been undertaken by Linden et al (subm.), who show that they are in very good agreement within the error bars.   Therefore, any differences we find in the spatial properties of eYSCs and optical YSCs during our analysis are not attributable to how the physical properties were derived. 



\section{Method}\label{sect_method}

\subsection{INDICATE}

To describe the spatial distribution of the stellar cluster population we use the 2+D INdex to Define Inherent Clustering And Tendencies (INDICATE) tool introduced by \citet{2019A&A...622A.184B}. INDICATE belongs to a class of statistical descriptors known as Local Indicators of Spatial Association (LISA; \citealt{Anselin_LISA}), which are defined as characterising discrete distributions by calculating a parameter value for every unique point, rather than a single parameter for the collective whole. Hereafter, we use `association' strictly in the statistical sense of \emph{spatial association}, and not to denote OB-type stellar groupings such as the Scorpius-Centaurus association.

INDICATE quantifies the degree of association of each and every cluster in our catalogue. No a priori knowledge is required, nor assumptions made about, any sub-structure present in the dataset. Extensive statistical testing by \citet{2019A&A...622A.184B} showed it to be robust against outliers and edge effects, as well as independent of a dataset’s size, shape and density. \citet{ 2022A&A...659A..72B} demonstrated INDICATE is robust against incompleteness up to 83.3$\%$, whereas \citet{2024MNRAS.527.5448B} showed that conclusions about the spatial properties of a dataset are unchanged even when there are few true members ($<\,2\%$) or significant 
($>20\%$) contamination by non-members that have been mistaken as true members of the dataset. INDICATE can be applied to any discrete astronomical dataset, observed or simulated, on any scale, and the results obtained for datasets of different densities, size, shape, and parameter space can be directly compared. Previous applications include describing spatial behaviours of: observed Young Stellar Objects as a function of evolutionary stage in NGC\,2264 \citep{ 2020A&A...636A..80B}; stars formed from a cloud-cloud collision simulation over a period of 1\,Myr \citep{2022MNRAS.514.4087B}; the observed stellar population of Orion and their potential evolution over the next 1\,Myr (simulated; \citealt{2023MNRAS.519.4142S}); mass segregation across various individual star clusters and forming complexes (\citealt{2022MNRAS.510.2864B}, \citealt{2019A&A...622A.184B}); differences in the degree of spatial clustering of the stellar component of the bar, inner spiral arm, outer arm, and inter-arm region of a simulated Milky-Way type galaxy \citep{2023MNRAS.524..555A}.   

Here we use INDICATE to examine how the spatial behaviour of NGC\,628’s clusters varies with age, mass and galactic position. The tool works by examining the local neighbourhood of a stellar cluster, comparing the number of clusters found there to that expected if it was in a random distribution, and assigning an index to denote whether the cluster is spatially concentrated – and if so, quantifies the degree of concentration.

The full details on INDICATE can be found in \citet{2019A&A...622A.184B}, but we provide here a summary of how the technique works in practice.
The first step is to convert the catalogue's J2000 sky coordinates to a local X-Y coordinate frame to ensure all dimensions have the same scaling. The number density of the catalogue is then calculated by dividing the total number of stellar clusters, $N_{tot}$, by the total 2D surface area of the parameter space. Next, an evenly spaced uniform (i.e., definitively non-clustered) control distribution with the same number density across the parameter space is generated. For each stellar cluster $j$ the Euclidean distance, $r_j$, to its $5^{\rm{th}}$ nearest neighbour in the control distribution is measured:
 
\begin{equation}
\\ r_j=\sqrt{(x_j-x^{\text{con}}_{5})^2 +(y_j-y^{\text{con}}_{5})^2}
\end{equation}

\noindent where $(x_j,y_j)$ and $(x^{\text{con}}_{5},y^{\text{con}}_{5})$ are the respective coordinates of the stellar cluster and $5^{\rm{th}}$ nearest neighbour in the control distribution. The average for the catalogue, $\bar{r}$, is calculated as: 

\begin{equation}
\\ \bar{r}=\frac{\sum\limits_{j=1}^{N_{tot}} r_j}{N_{tot}}
\end{equation}

\noindent Finally, the actual number of stellar clusters within a radius $\bar{r}$ to stellar cluster $j$, $N_{\bar{r}}$, is then counted and the index, $I_{j,5}$, calculated:

\begin{equation}
\\ I_{j,5}=\frac{N_{\bar{r}}}{5}
\end{equation}

\noindent where $I_{j,5}$ is unit-less and has a value in the range of $0\le I_{j,5}\le \frac{N_{tot}-1}{5}$. Significant values are identified by calibrating the index with 100 random distributions of size $N_{tot}$ in the parameter space of the catalogue to determine the ‘significance threshold’, $I_{sig}$, defined as:

\begin{equation}\label{eq_Isig}
\\ I_{sig}=\bar{I}^{\text{random}}_{5}+3\sigma
\end{equation}

\noindent where $\bar{I}^{\text{random}}_{5}$ is the mean index derived for the points of the random distribution and $\sigma$ the standard deviation. Using this definition, 99.7$\%$ of the index values derived for randomly distributed points will have $I^{\text{random}}_{5}< I_{sig}$. Values greater than the significance threshold denote spatial clustering, and for these, higher values denote greater degrees of spatial association. It is important to note that the index is not a descriptor of density, such that values derived for datasets of disparate densities can be directly compared, and will be unchanged if the spatial distributions are identical (see Section\,2.1.4. of \citealt{2019A&A...622A.184B} for a discussion). The minimum sample size for INDICATE is 50 datapoints \citep{2019A&A...622A.184B}. The use of a uniform control field (rather than one where we follow an underlying non-uniform distribution) is adequate for our purpose, as we are exploring how the spatial association of the stellar component vary across the galactic disc. In other words, we aim to assess clusters' non-random spatial relationships, then directly compare, quantify, and identify small-scale local variations in these. By changing the control field to a non-random distribution (for example, one that follows the underlying stellar distribution of NGC\,628) removes our ability to identify random spatial relationships and thus quantify and compare clusters' spatial associations across the galaxy.

\subsection{Accounting for the observational Age-Mass Bias}\label{sect_bias}

\begin{figure}
\includegraphics[width=0.45\textwidth]{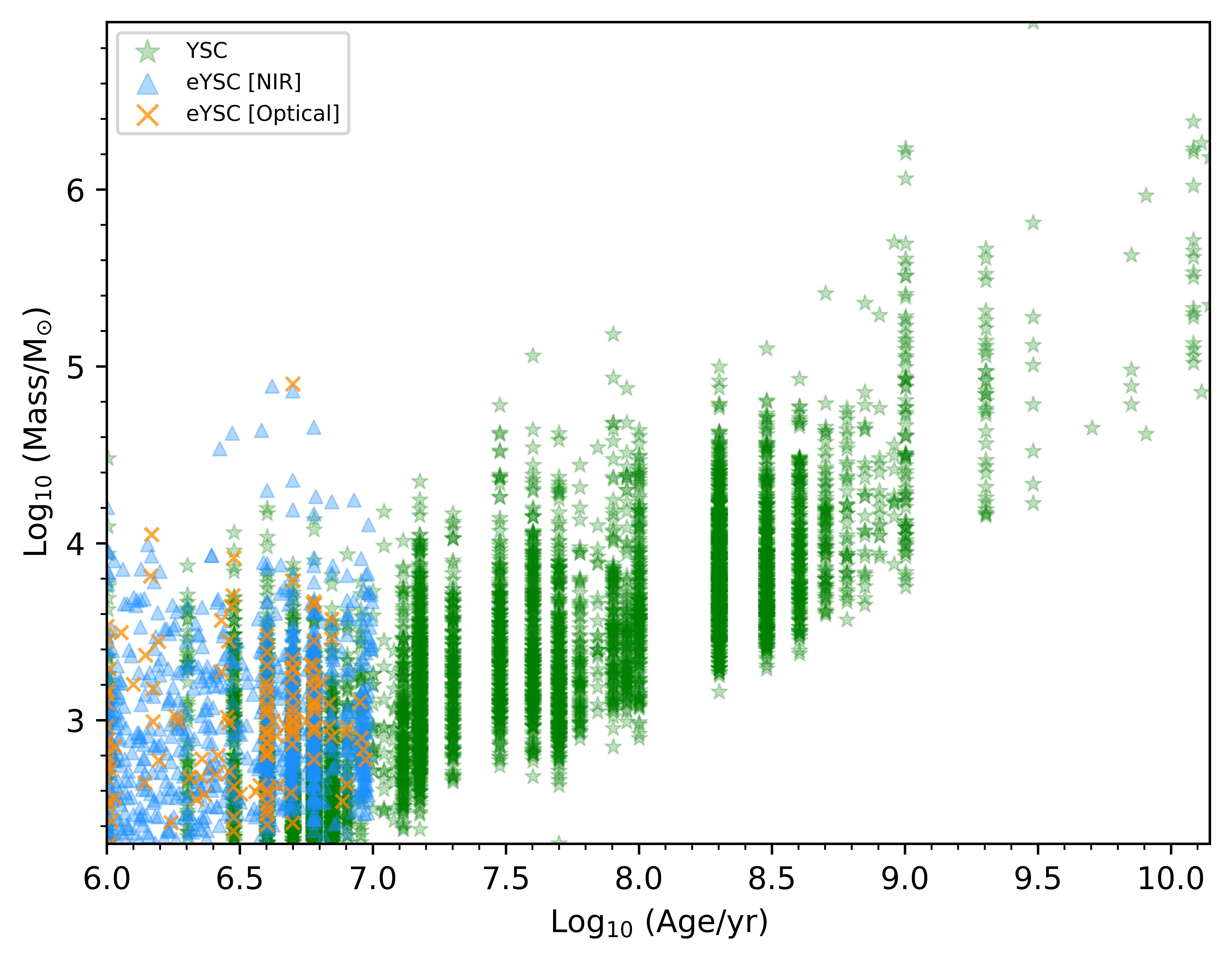}
\caption{Plot of age against mass for our cluster catalogue. Green stars represent optical YSCs, orange crosses eYSCs that are optically visible, and blue triangles eYSCs that are not optically visible. See Sect.\,\ref{sect_cat_selection} for details. \label{fig_cat_agemass}}
\end{figure}

As shown in Figure\,\ref{fig_cat_agemass}, there is an age-mass bias in our catalogue which is well known in the literature and is due to the limited depth of the data combined with the fading of star cluster light due to stellar evolution \citep[e.g.][among many others]{2017ApJ...841..131A}. This means that for clusters older than $10$ Myr, the minimum mass that we are able to detect increases steadily with age, producing a lower limit with a rising slope as a function of age. We note that no artificial spreading was applied to the figure at <10Myr: the vertical banding for the optical YSCs arises directly from the Yggdrasil spectral evolutionary models, whereas the eYSC ages were derived using CIGALE (see Sect.\,\ref{sect_cigale}\,and\,\ref{sect_ygg} for details), and the logarithmic age axis can additionally enhance the visual prominence of such discrete loci.

At all ages, we also see that the number of massive clusters is steadily decreasing due to the shape of the cluster mass function: a power law with slope $-2$, implies that it is 10 times more likely to sample a cluster of $10^3$~M$_\odot$ than one at $10^4$~M$_\odot$ and so on. 

This, combined with the age binning of the sample, produces a visual increasing upper mass limit with age: since the x-axis on Fig.\,\ref{fig_cat_agemass} is in logarithmic scale, the linear-age interval increases to the right in the diagram. Therefore even if NGC\,628’s star/cluster formation rate remains constant with time, the total number of clusters that form in each equal interval of log(Age) increases. Thus, the likelihood of a very high mass cluster being within a given log(Age)-bin increases (see e.g., \citealt{2003AJ....126.1836H}). 

We note that eYSCs densely populate the age bins $<$ 10 Myr. This population of recently formed (and still partially embedded) star clusters has largely been missed by HST surveys due to the lack of IR coverage, which is less affected by dust attenuation. Because of the clustering nature of star formation, eYSCs naturally include associations that are not gravitationally bound and dissolve on time scales of 10 Myr. 
For example, we know from observational studies of our own Galaxy between $93-96\%$ of eYSCs will not survive the emergence phase to become bound star clusters (\citealt{2003ARA&A..41...57L}, \citealt{2010MNRAS.405..857G}), and with a value of $\sim$ 85-90\% found for the nearby spiral galaxy NGC\,6946 \citep{2010MNRAS.405..857G}, it is not unreasonable to anticipate a value similar to these for NGC\,628. By comparing the number of eYSCs and those that are classified as likely bound in the optical YSCs catalogues (e.g, class 1 and 2), Adamo et al (subm.) find that only $\sim$ 20\% of eYSCs are expected to survive within 10 Myr in NGC\,628. That is, it is expected that the eYSC population would be more densely populated than the recently emerged YSC population. Moreover, only higher mass YSCs are expected to survive to ages of 100\,Myr+, due to mass loss from internal and external disruption events, so we would not expect to see a large low-mass population in older age bins (\citealt{2003MNRAS.338..717B},  \citealt{2010IAUS..266...69G}, \citealt{2019ARA&A..57..227K}). 

Despite these issues, INDICATE is, by design, robust against both uniform and patchy incompleteness in datasets. Extensive testing has shown that when it is applied to the full and incomplete versions of a dataset, only small variations between the index values occurs (for an incompleteness of up to $83.3\%$) and there is no change in the conclusions regarding the spatial behaviours of a dataset \citep{2022A&A...659A..72B,2024MNRAS.527.5448B}. As a check, we compared the index derived for the full sample and with various mass and age cuts, which are presented in Appendix\,\ref{sect_app1} concluding that results do not change significantly when cuts are applied. Therefore, for the purpose of this study, we do not need to apply age or mass cuts to the cluster sample to compensate for the age-mass completeness bias. One exception is that, for the study of the spatial behaviours for different mass clusters (Sect.\,\ref{sect_results_mass}), we apply a mass-age cut in order to minimise the potential propagation of an evolutionary trend onto the mass trend.


\section{Spatial Behaviour of Clusters in NGC\,628}\label{sect_results}

\begin{figure*}
\includegraphics[width=\textwidth]{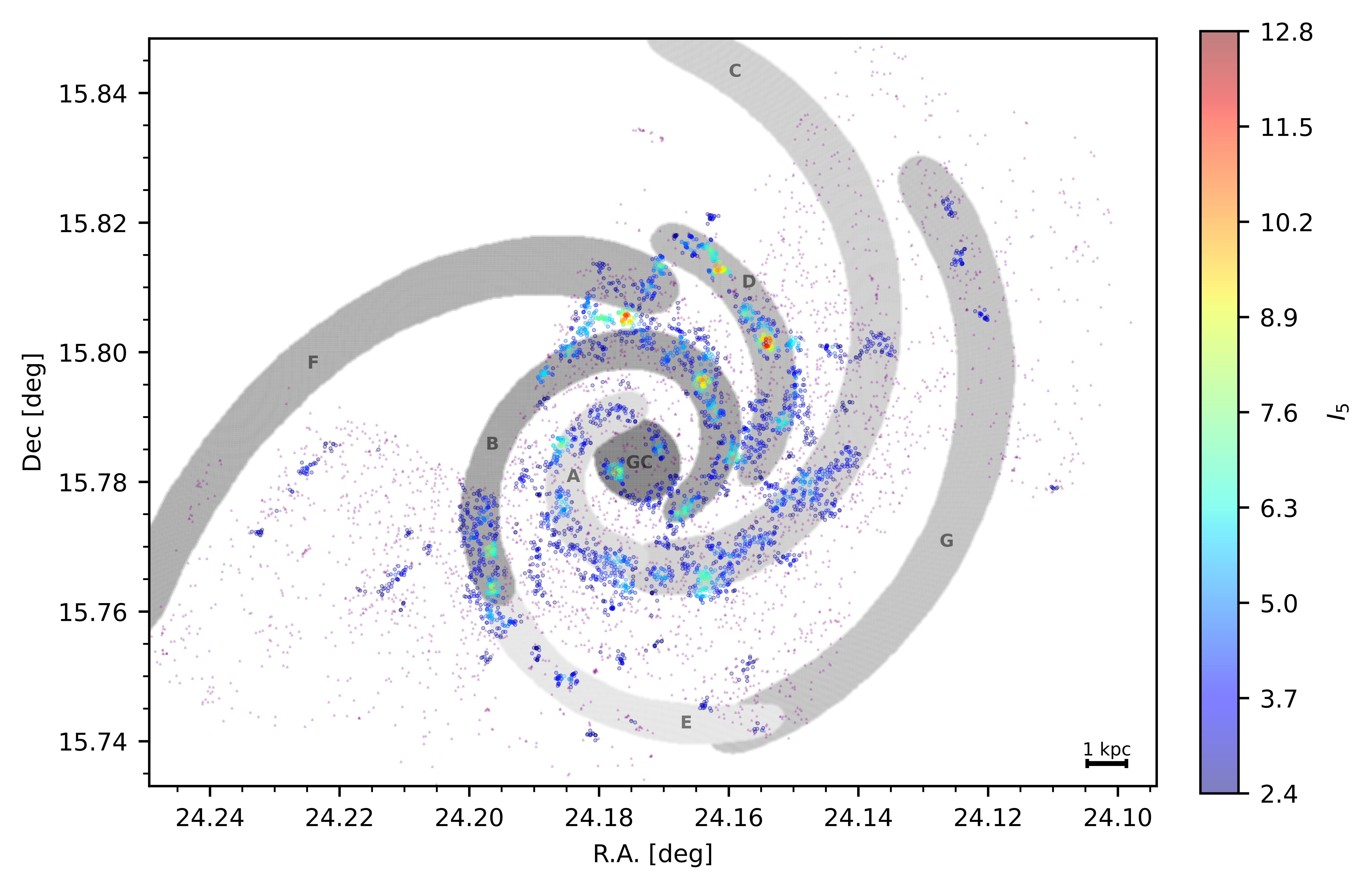} \\
\caption{Plots of index values, $I_5$ , calculated by INDICATE for NGC\,628, with the environmental mask by \citet{2021A&A...656A.133Q} overlaid showing the positions of Spiral Arms A-G (marked by various shades of grey), and Galactic Center (dark grey circle, 'GC'). Clusters with an index value above the significance threshold ($I_5 > 2.4$) are represented by circle markers, colour-coded by their $I_5$ index, where higher values denote greater degrees of spatial concentration. Purple triangles are clusters with an index value below the significance threshold. \label{fig_RaDec_I5}}
\end{figure*}

\subsection{Galactic Regions}\label{sect_results_gp}

INDICATE was applied to the combined sample of stellar clusters, and index values were derived for each of the 6890 clusters, as shown in Figure\,\ref{fig_RaDec_I5}. Table\,\ref{tab_galpos} gives statistics on the index values derived for each region of NGC\,628. The significance threshold was determined using Eq.\,\ref{eq_Isig} to be $I_{sig}=2.4$.

Only $51.8\,\%$ of the clusters in the galaxy are spatially concentrated, with a median value of $\tilde{I}^{C}_5=3.8$. The proportion of the population in each region which is spatially dispersed varies, with the lowest proportion in the spiral arms ($29.7\,\%$) and galactic centre ($25.7\,\%$), and the largest proportion in the interarm regions ($69.4\,\%$). Interestingly, the degree of association of the concentrated population also varies by region, with the spiral arms showing the greatest (median index $\tilde{I}^{C}_5=4.0$), followed by the galactic centre ($\tilde{I}^{C}_5=3.8$), and the interarm regions typically demonstrating the lowest ($\tilde{I}^{C}_5=3.2$). 


Figure \ref{fig_RaDec_I5} shows that all spiral arms host strongly spatially concentrated cluster populations, with typical INDICATE index values above the galaxy-wide average. However, the clusters’ spatial behaviour varies both between the spiral arms and along the length of each arm. Arm D hosts both the greatest proportion of spatially concentrated clusters ($83.3\%$) and typical degree of association ($\tilde{I}^{C}_5=4.8$). Two areas were found to have particularly high degrees of association, with index values here being the greatest found in the galaxy with a maximum value of $I_{5}=12.8$ (i.e. in the local neighbourhood of the clusters there are 5.3 times the number of clusters above random expectation, and 3.4 times more than typical for the non-dispersed cluster population in NGC\,628). Arm B hosts the second greatest proportion of spatially concentrated clusters ($81.9\%$) and typical degree of association ($\tilde{I}^{C}_5=4.4$). Like in D, there are a few areas of high spatial concentration (albeit slightly lower intensity, max. $I_{5}=10.4$), with a high-intensity area found in between the designated bounds of B and F. The remaining spiral arm segments (for which we have coverage) contain a few areas of higher degrees of association of similar intensity, but the typical association here is comparable to that found for the general concentrated population in NGC\,628 and the proportion of spatially concentrated clusters is significantly lower than that found in D and B ($60.5-74.8\%$).

To assess whether these differences are statistically significant, we applied two-sample Kolmogorov-Smirnov (K-S) tests, which evaluate whether two samples are drawn from the same parent distribution and therefore test for differences in their overall distributions (significance threshold $p<0.05$). We found that clusters in the spiral arms and galactic centre have a similar underlying spatial distribution ($p > 0.05$), but clusters in the interarm region have significantly different spatial behaviour ($p<0.05$). Clusters in spiral arms  B and D have distinct clustering tendencies ($p<0.05$), but A, C and E have similar clustering tendencies ($p>0.05$). This result confirms that the spatial behavior of clusters varies with galactic position. We explore the possibility of variations with evolutionary stage and mass in the next sections.

\begin{table}
\begin{center}
\caption{Table detailing for each region: total number of clusters present, the percentage of clusters found to be spatially concentrated (i.e. $I_5>I_{sig}$) and their median index value ($\tilde{I}^{C}_5$)\label{tab_galpos}}
\label{demo-table}
\begin{tabular}{||c c c c ||} 
 \hline
Region & Total clusters & $\%$ Concentrated & $\tilde{I}^{C}_5$  \\ 
 \hline\hline
All & 6890 & 51.8 & 3.8 \\
 \hline\hline
GC & 206 & 74.3 & 3.8\\
 \hline
Inter Arms & 3224 & 30.6 & 3.2\\
 \hline
Spiral Arms & 3460 & 70.3 & 4.0\\
 \hline
  \hline
SA-A  & 409 & 74.8 & 3.8\\
 \hline
 SA-B  & 932 & 81.9	& 4.4 \\
\hline
 SA-C  & 868 & 63.8 & 3.8\\
\hline
 SA-D  & 599 & 83.3 &  4.8\\
\hline
 SA-E  & 223 & 60.5	& 3.6\\
\hline

\end{tabular}
\end{center}
\end{table}


\subsection{Evolutionary Stage}\label{sect_results_age}

\begin{figure*}
\includegraphics[width=0.5\textwidth]{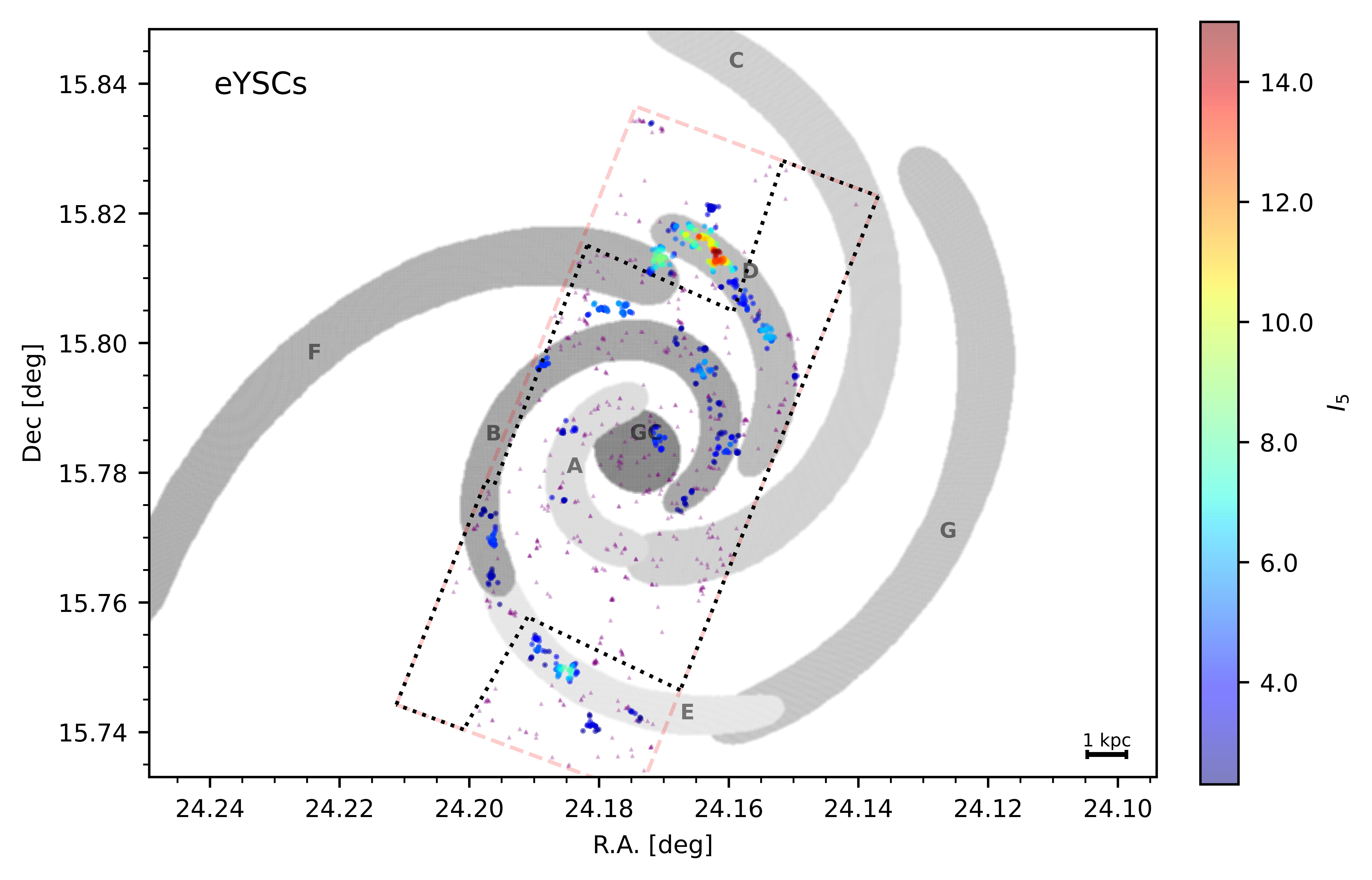}
\includegraphics[width=0.41616\textwidth]{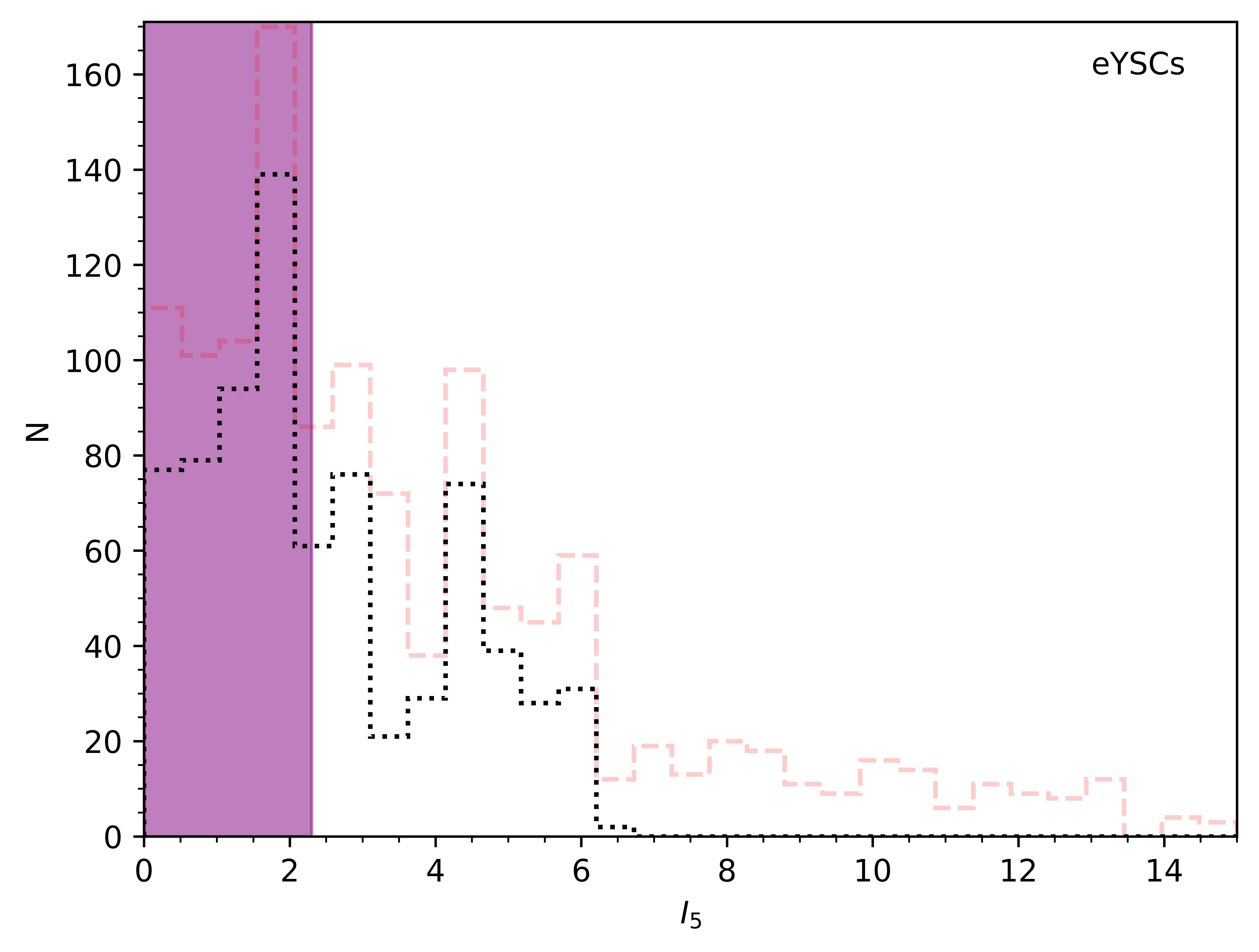}
\includegraphics[width=0.5\textwidth]{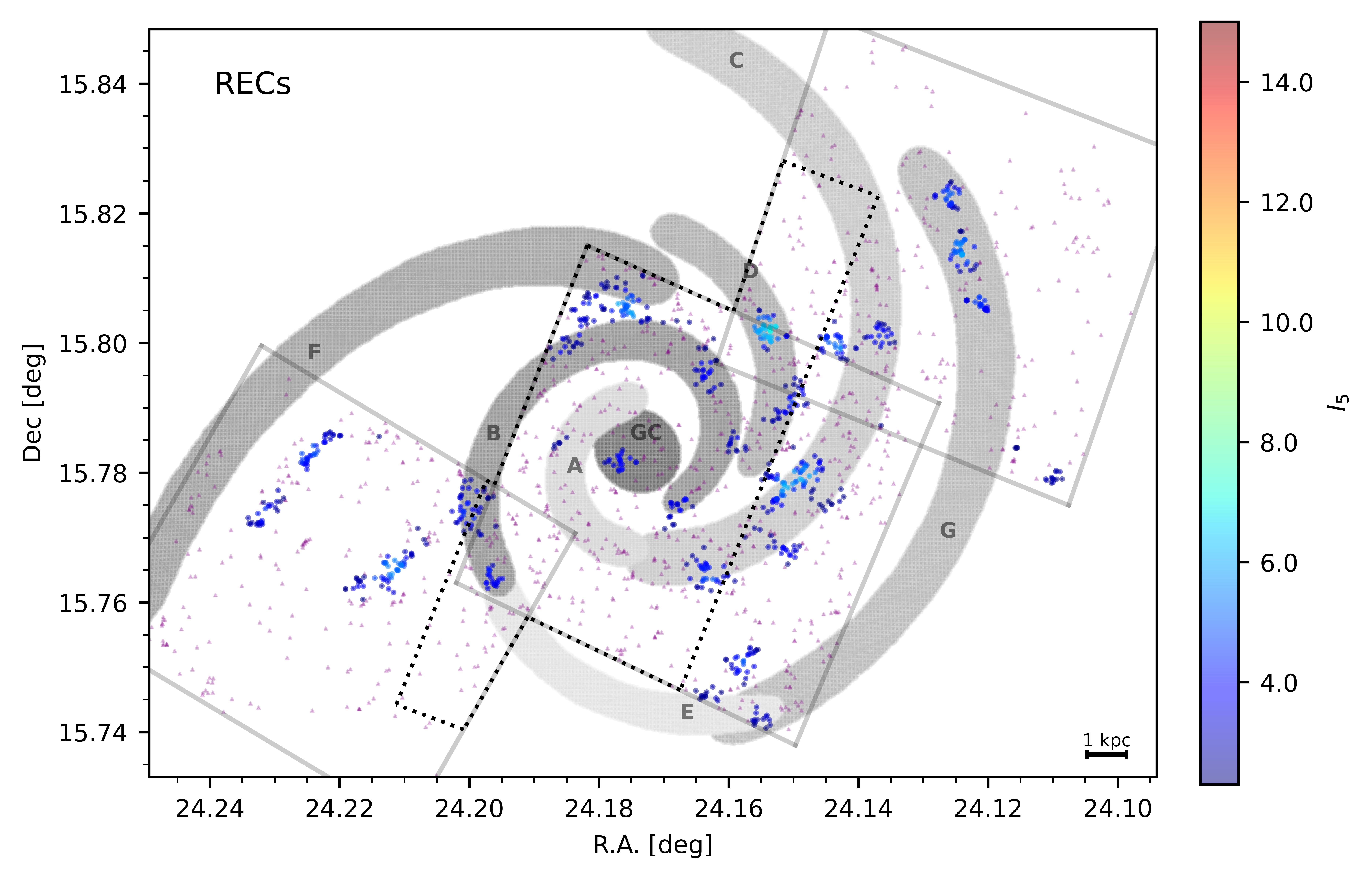}
\includegraphics[width=0.416\textwidth]{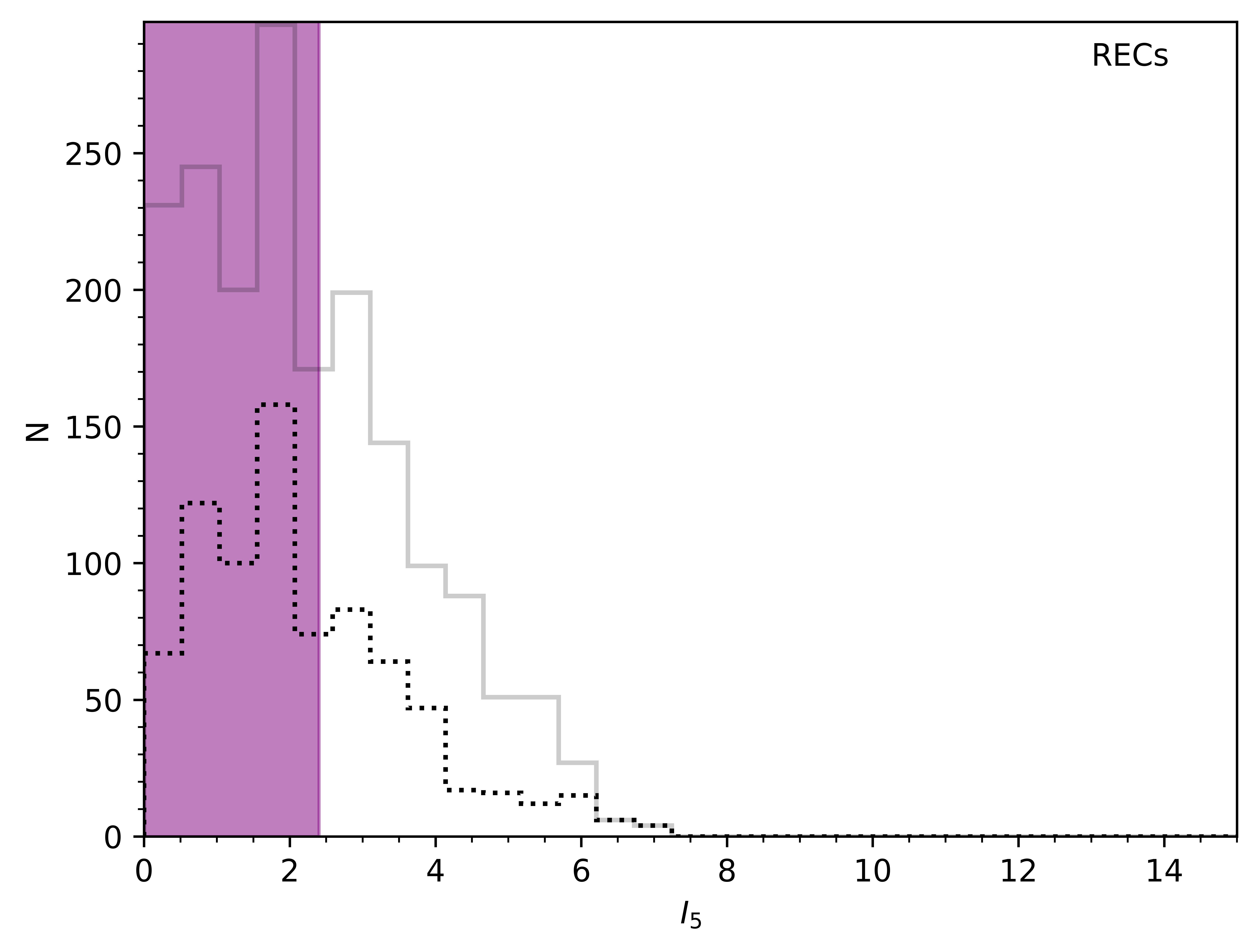}
\caption{Index values, $I_5$, plotted as a function of galactic position (left) and as a histogram (right), for the (Top:) eYSCs and (Bottom:) recently emerged clusters (RECs), defined as the optical YSCs $<10$\,Myr, discussed in Sect.\,\ref{sect_results_eYSCs}. The index values represent the degree of association clusters have with their counterparts of similar age. On the left panels, clusters with an index value above their sample's significance threshold are represented by circle markers, colour-coded by their $I_5$ value, as per the respective colour bars. Clusters with index values below the sample's significance threshold are marked by Purple triangles on the left panels, and shaded areas on the histograms. On the left panels, the environmental mask by \citet{2021A&A...656A.133Q} is overlaid to show the location of Spiral Arms A-G (marked by various shades of grey), and Galactic Center (dark grey circle, labelled 'GC'). The grey, red-dashed and black-dotted boxes denote the position of the HST mosaics, the JWST coverage, and their overlap region respectively. Line colour and style in the histograms on the right correspond to that of clusters within the respective coverage and overlap areas, as denoted in the position plots. \label{fig_I5_eYSCs}}
\end{figure*}


\subsubsection{Emerging YSCs versus optical YSCs ($\le10$\,Myr)}\label{sect_results_eYSCs}

Here we explore the differences in the spatial distributions of the emerging young clusters still (partially) embedded in their natal clouds (eYSCs) and the non-embedded optical YSCs ($\le10$\,Myr) which we will refer to in this section as Recently Emerged Clusters (RECs). We define RECs as clusters that are $\le10$\,Myr and not in the eYSC catalogue. For consistency in cross–sample comparisons, we cap the RECs at a maximum age of \(10\,\mathrm{Myr}\) when comparing to eYSCs, matching the eYSC catalogue’s upper age limit. Given that both sub-samples span the same ages, differences on their relative spatial distribution could be indicative of the spatial evolution of the clusters as they emerge from their natal clouds. Unfortunately, as the eYSCs were identified from the JWST image, and the RECs from the HST images, there is only a relatively small overlap area containing both samples as shown on Figure\,\ref{fig_I5_eYSCs}. Therefore, we restrict our comparison to clusters inside this area (as, for example, it is unknown whether RECs have similarly high degrees of association as found for the eYSCs towards the northern end of Arm D). This overlap region contains 750 eYSCs and 785 optical RECs. Their respective age and mass distributions are shown in Fig.\,\ref{fig_hist_embedded}.

We find that the proportion of eYSCs that are concentrated ($42.9\%$) is marginally higher than that of the REC population ($38.6\%$), and that eYSCs also are typically a factor of $\sim\,1.3$ times more tightly associated with each other than their emerged counterparts. From the left panels of Fig.\,\ref{fig_I5_eYSCs} the galactic distribution of the two concentrated populations share some commonality, as well as some differences in the specific areas of high and low degrees of association which vary across the disk. Plotting a histogram of the indices (right panels, Fig.\,\ref{fig_I5_eYSCs}), both distributions have a similar range on index values, especially within the overlap region, but the shape of the distributions is different (with one smoothly decreasing, while the other presents strong troughs and peaks). To determine the significance of these disparities, we compare the cumulative distribution functions of the index values of (i) all and (ii) only above the significance threshold, for all clusters in each group inside the overlap area using a 2 sample K-S test, using a significance boundary of $p<0.05$. Finding $p<<0.05$ in both tests, we conclude eYSCs and RECs ($\le10$\,Myr) do not have the same underlying spatial distribution. This suggests that there is a distinct change in the spatial behaviour of clusters as they emerge from the natal clouds, with them typically becoming significantly less tightly associated than when they are still (partially) embedded. This could potentially be due to clusters moving away from their birth place as they clear their natal cloud (and therefore moving away from other nearby young clusters), or even due to the fact that some clusters may not survive once the natal cloud has been cleared (as they lose the potential well from the gas), thus dispersing their stellar population and `disappearing' from the population of recently emerged clusters. Such rapid early dissolution (\textit{`infant mortality'}) and its environmental dependence have been both theoretically predicted and observationally observed in other system (e.g. \citealt{2010ApJ...712..604E};\citealt{Bastian2012}; \citealt{Krumholz2019}; \citealt{2022ApJ...935..166L}).

\subsubsection{Age}\label{sect_results_evo}

\begin{figure*}
\includegraphics[width=0.5\textwidth]{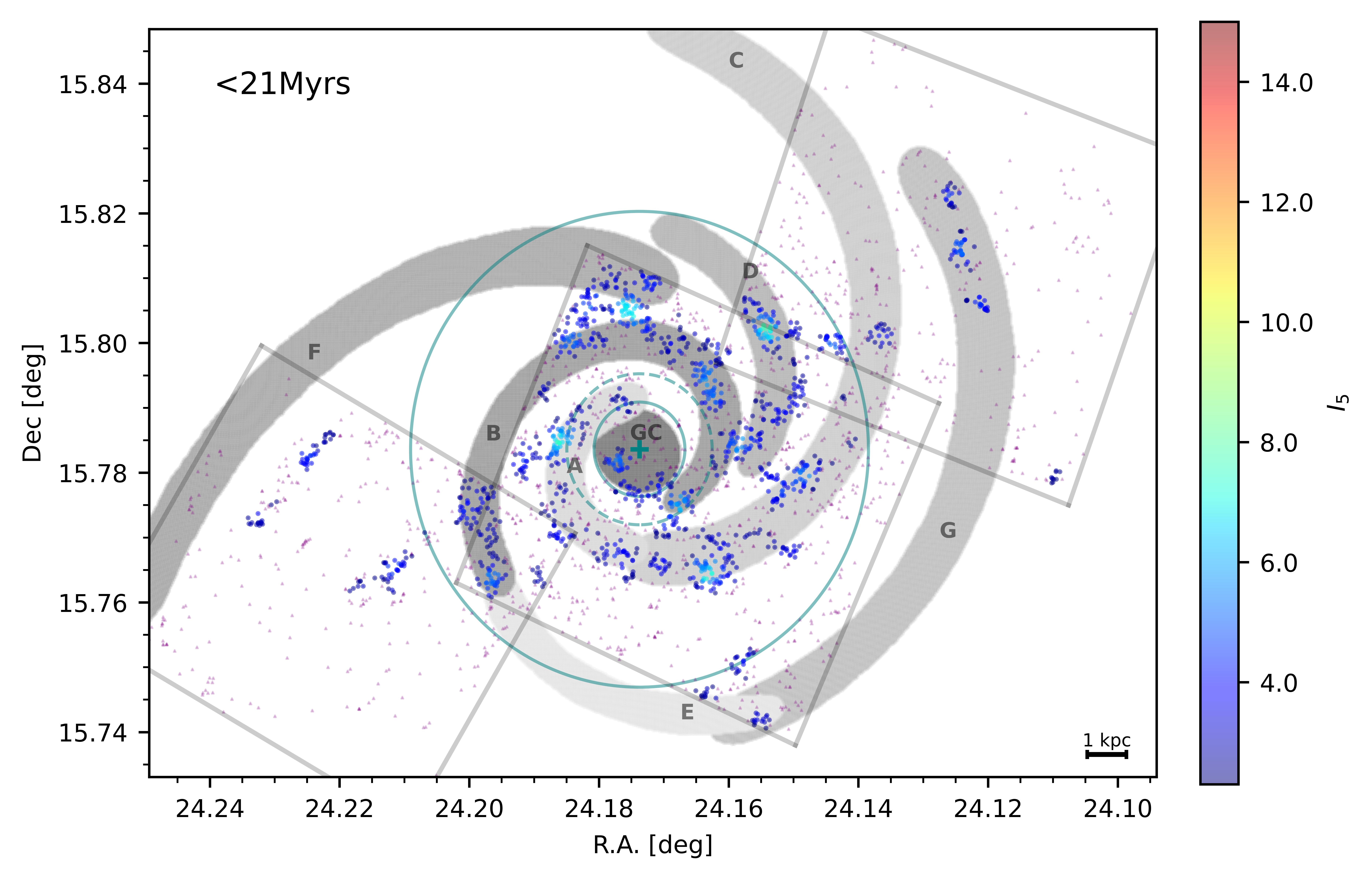}
\includegraphics[width=0.416\textwidth]{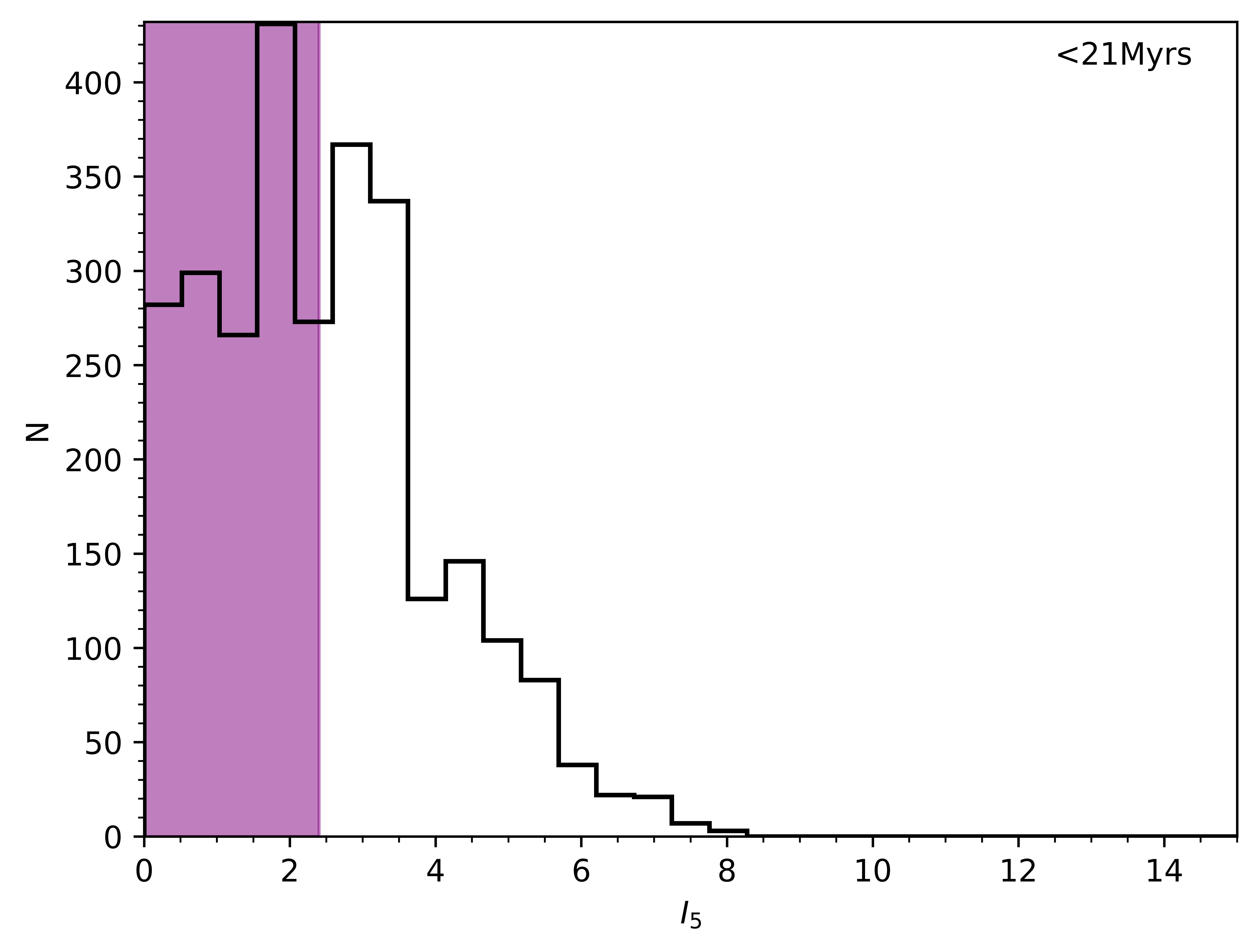}
\includegraphics[width=0.5\textwidth]{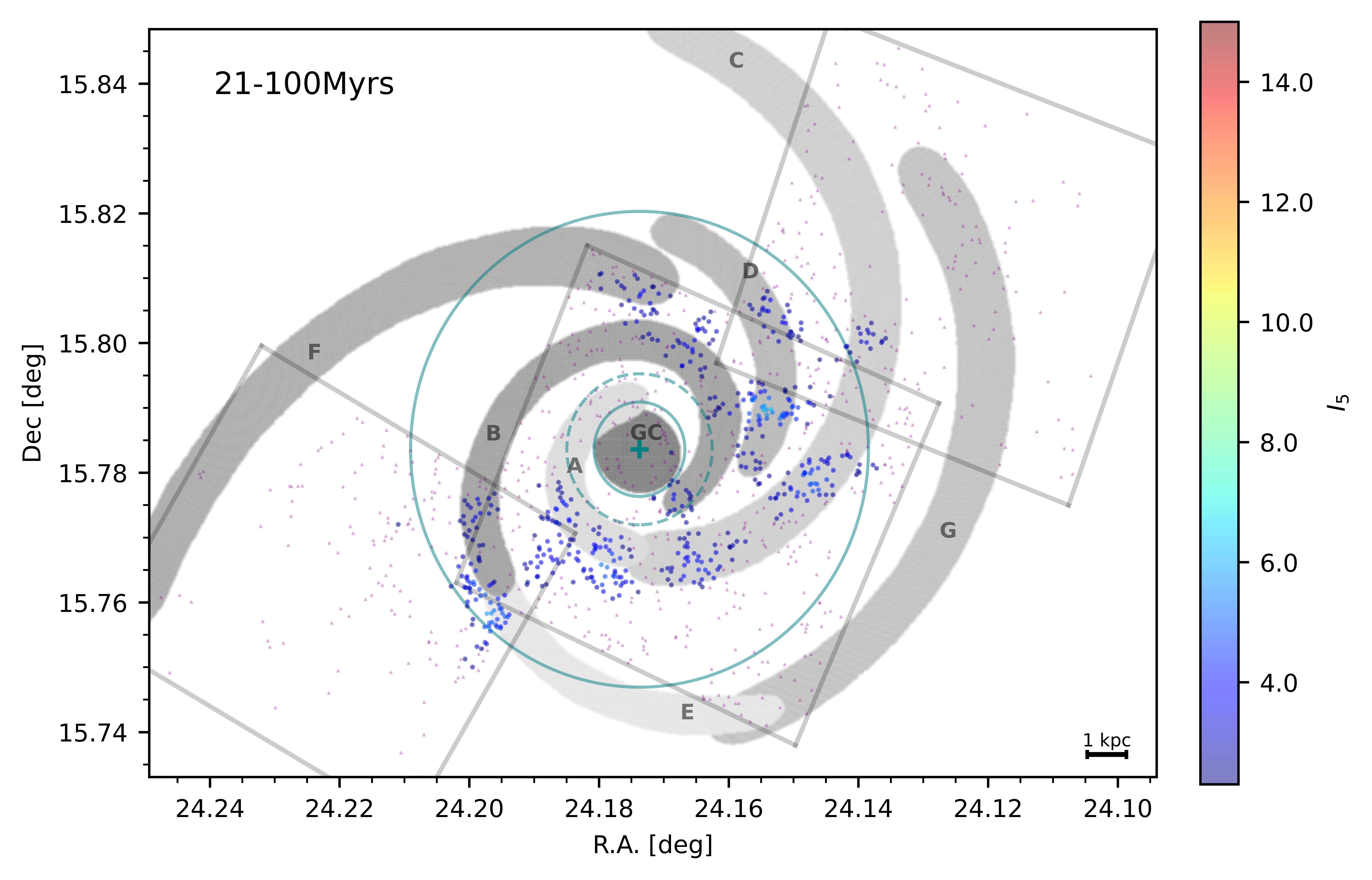}
\includegraphics[width=0.416\textwidth]{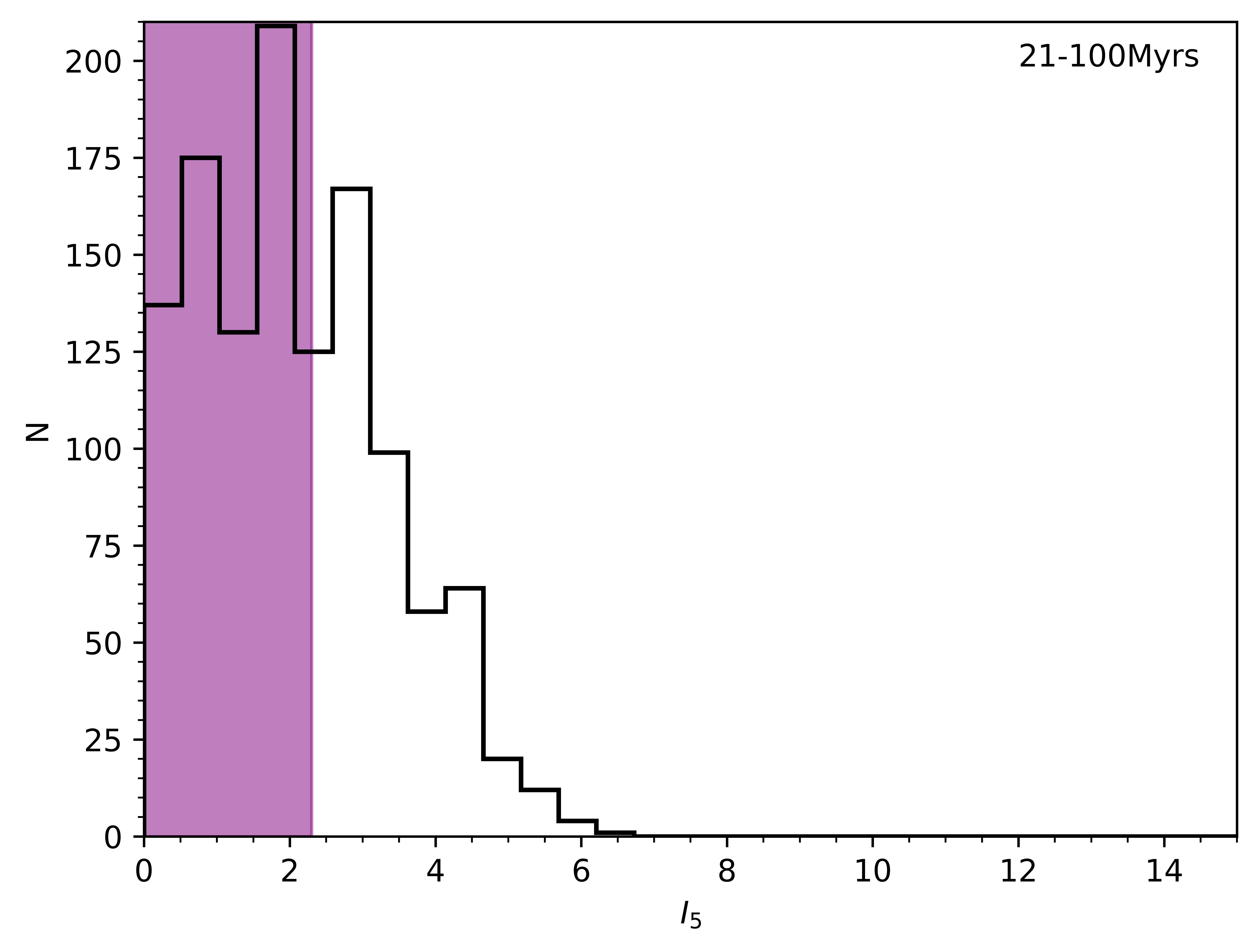}
\includegraphics[width=0.5\textwidth]{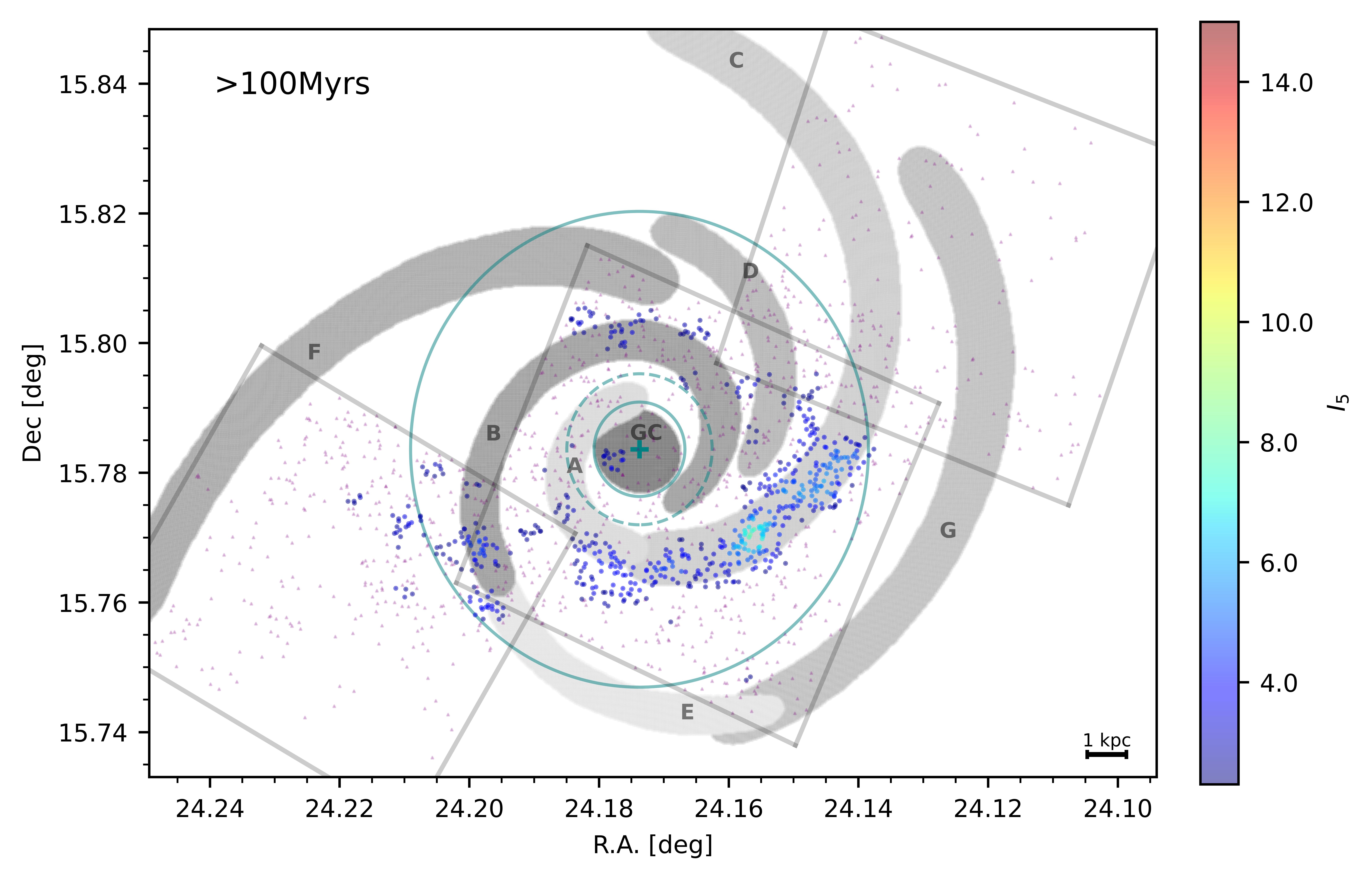}
\includegraphics[width=0.416\textwidth]{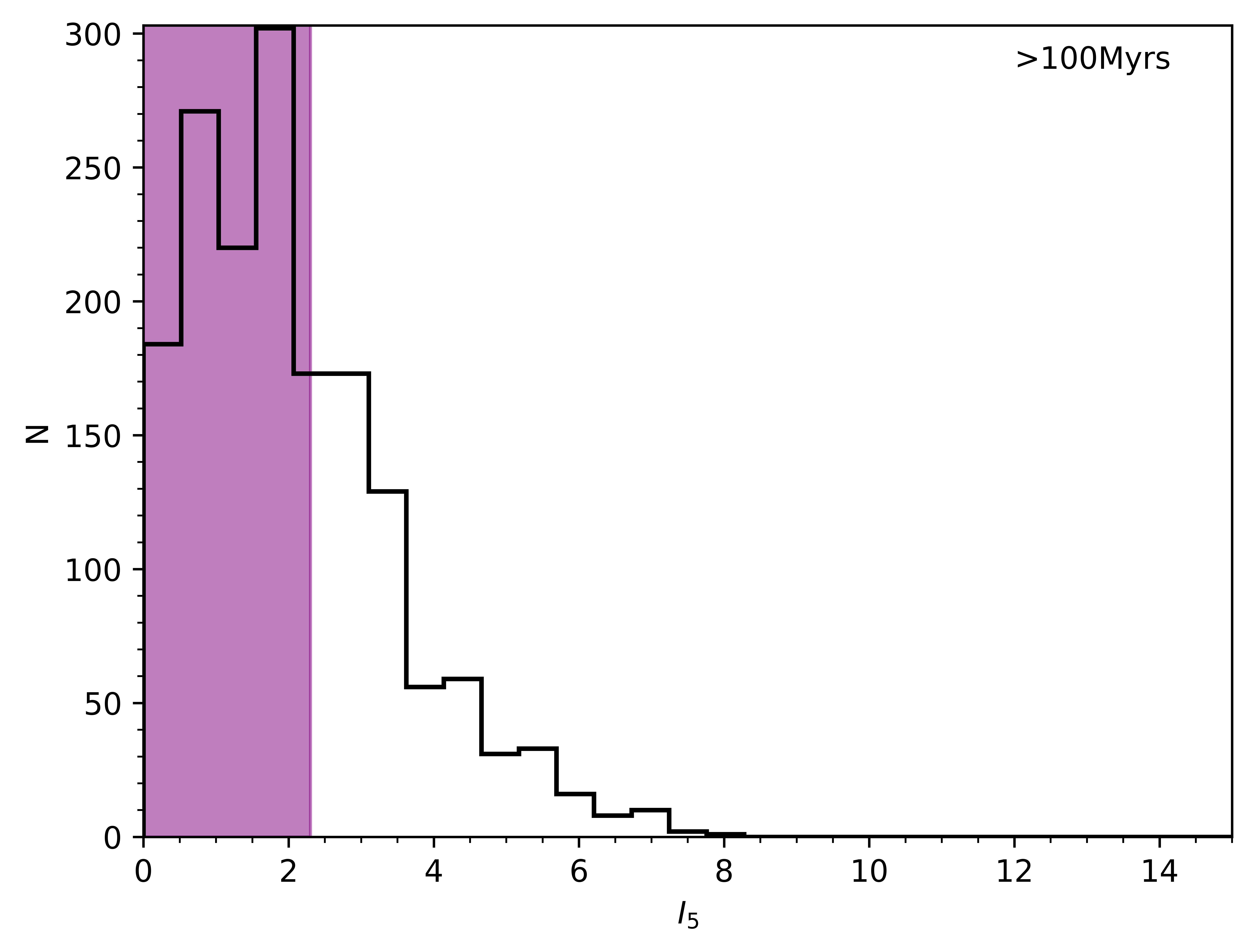}
\caption{Index values, $I_5$, for the Young (top), Intermediate (middle) and Old clusters (bottom), as discussed in Sect.\,\ref{sect_results_evo}. The index values represent the degree of association clusters have with their counterparts of similar age. Left panels show $I_5$ as a function of galactic position, and right panels as a histogram. The grey boxes on the left panel denote the position of the HST coverage, with the environmental mask by \citet{2021A&A...656A.133Q} overlaid. The teal cross and ellipses mark the galactic center, a radius of 1.25\,kpc, 2\,kpc and the galactic co-rotation radius (6.3\,kpc; \citealt{2021MNRAS.508..912S}), respectively. Clusters with an index value above their sample's significance threshold are represented by circle markers, colour-coded by their index value. Clusters with index values below the sample's significance threshold are marked by Purple triangles and shaded on the position and histogram plots. \label{fig_RaDec_I5_age}}
\end{figure*}

\begin{table}
\begin{center}
\caption{Table detailing the spatial behaviour of clusters by evolutionary stage. Shows the total number of clusters, percentage found to be spatially concentrated with those of a similar age, and their respective median index values. Stages are as defined in Sect.\,\ref{sect_results_evo}. Bracketed values represent those within the HST/JWST overlap area shown in Fig.\ref{fig_I5_eYSCs} \label{tab_age}}

\begin{tabular}{|c||c|c|c|c|}
 \hline
 Stage& Age & Total Clusters  & $\%$ Concentrated  & $\bar{I}^{C}_5$ \\
 \hline\hline
eYSCs & $\le$10\,Myr & 1216 & 55.1	& 4.8  \\
 & & (750) & (42.9)	& (4.4)  \\		
RECs & $\le$10\,Myr & 1813 & 41.5	& 3.4  \\
 & & (785) & (38.6)	& (3.4)  \\	
Young & $<$21\,Myr& 2805 &  49.4 & 3.4  \\
Intermediate & 21-100\,Myr & 1201  & 41.2 & 3.2 \\
Old &$>100$\,Myr &  1668 & 35.6 & 3.4 \\
 \hline
 \end{tabular}
\end{center}
\end{table}

In order to investigate the spatial behaviour of stellar clusters as a function of age (and including before and after the emerging phase), we consider our eYSC sample as per Sect.\ref{sect_results_eYSCs}, and we divide the optical (emerged) clusters of our sample into three groups: Young ($<21$\,Myr), Intermediate (21-100\,Myr) and Old ($>100$\,Myr). The upper age limit of the Young sample was chosen to include all potentially unbound clusters with pre-main sequence stars that have not yet dissolved \citep{2012A&A...547A.107S}. INDICATE was run on each of these sub-samples to ascertain the tendencies of clusters at a similar evolutionary stage to spatially associate with each other, and the strength of these associations. 

Figure\,\ref{fig_RaDec_I5_age} shows the distribution of the index values derived for each sample across the galactic disc, and Table\,\ref{tab_age} contains a summary of the statistics. At all evolutionary stages the concentrated population is primarily found across the spiral arms, but there are marked differences in their locations. As expected from the analysis of the distribution of RECs in Sect.\ref{sect_results_eYSCs}, the Young optical clusters (which here include up to 20\,Myr-old clusters) display areas of high degrees of association in Arms D, B and B/F as indicated in the top panel of Fig.\ref{fig_RaDec_I5_age}, which are somewhat similar to the distribution of eYSCs (albeit with different levels of concentration). However, this Young population also has higher degrees of association in A and C that were not seen in the eYSC population. Intermediate clusters have areas of higher degrees of association in Arm E, and D (but more southernly than the Young/eYSC populations). Old clusters have just one area of intense association in Arm C which was not observed in either the young or intermediate populations.  

Regarding the proportion of spatially concentrated clusters within each age group, eYSCs have the greatest at $55.1\%$ across the entire sample. A steady decrease with increasing cluster age is then observed, culminating in only $35.6\%$ of Old clusters across the entire disc found to be spatially concentrated. The intensity of these concentrations is also greatest for the eYSCs at $\tilde{I}^{C}_{5}=4.8$. Young, Intermediate and Old clusters are found to have similar intensities, at a factor of $\sim\,1.5$ times less that of the eYSCs. We check the significance of these findings by running 2 sample K-S tests on the index values for all clusters in each age group against those of the other age groups, with a significance boundary of $p < 0.05$. Subsequently, we reject the null hypotheses and verify that the exhibited spatial association behaviour of clusters at each evolutionary stage is distinct.

\begin{figure*}
\includegraphics[width=0.49\textwidth]{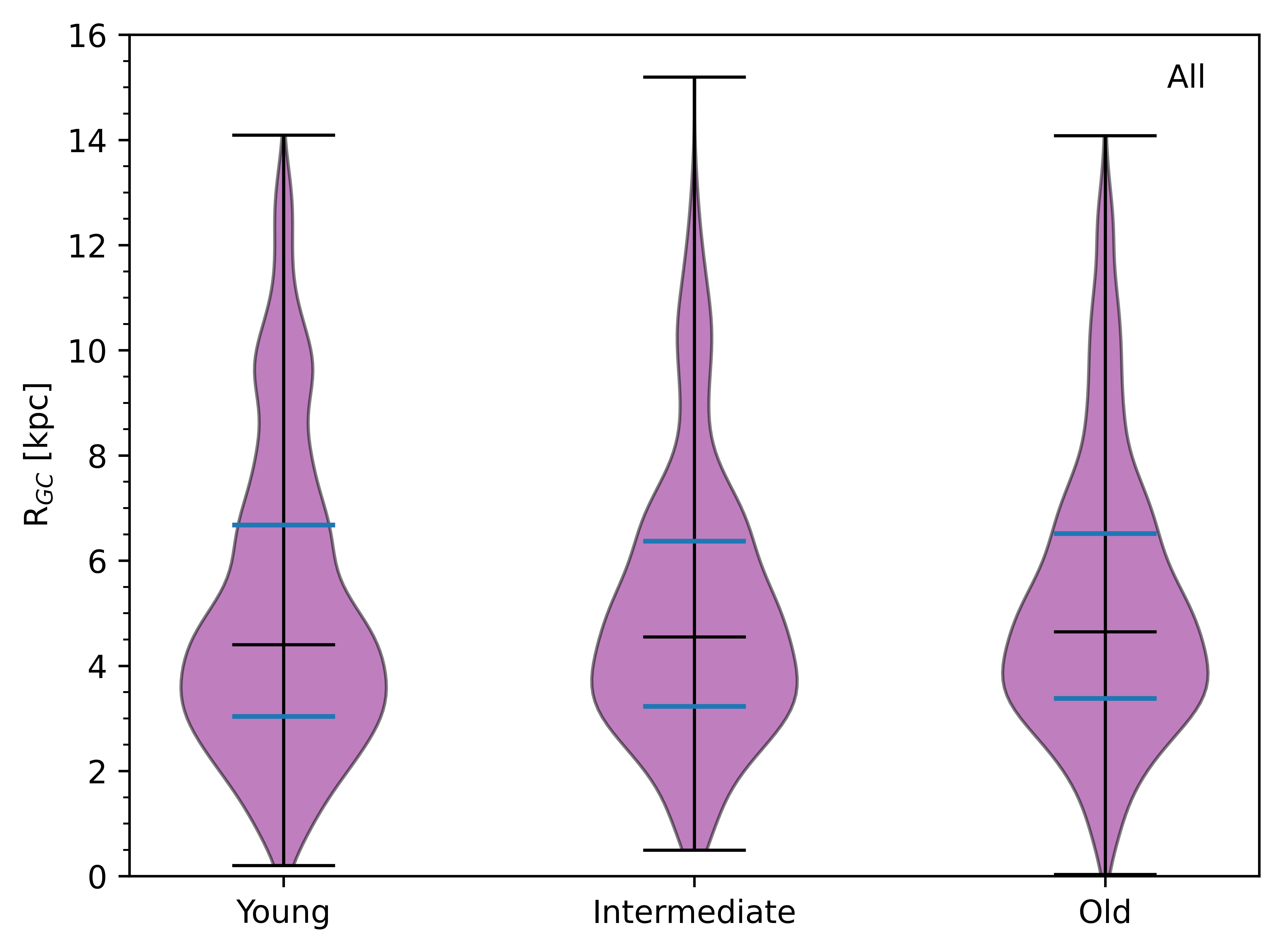}
\includegraphics[width=0.49\textwidth]{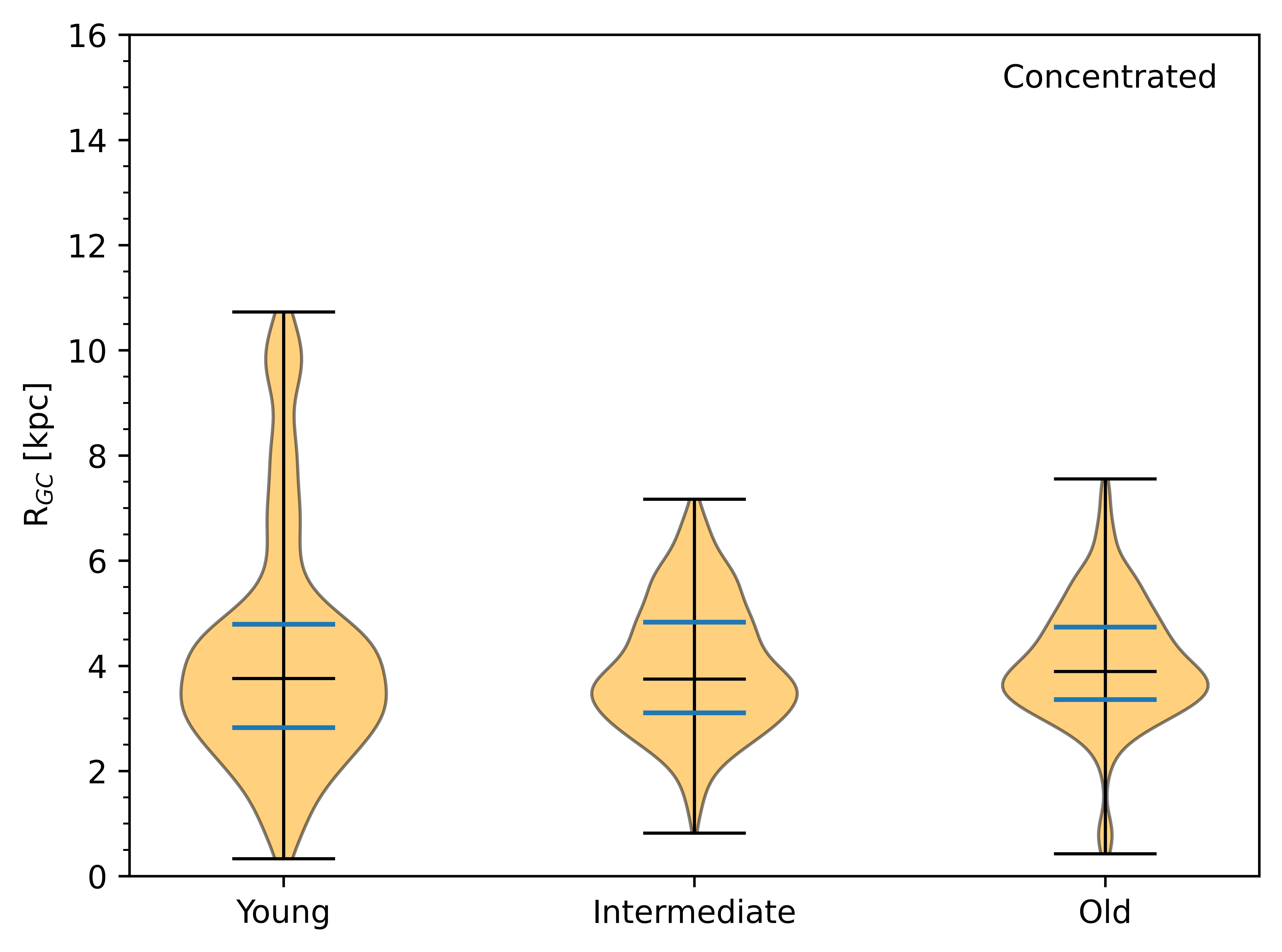}
\caption{Violin plots depicting the one-dimensional radial distributions of optical clusters for (Left:) all and (Right:) spatially concentrated of the Young, Intermediate and Old groups, as discussed in Sect.\,\ref{sect_results_evo}. The vertical extent of each violin corresponds to galactocentric radius, R$_{\rm{GC}}$, while the horizontal width represents the relative number density (probability density) of clusters at a given radius. Spatially concentrated clusters are defined as having an index value above their sample's significance threshold i.e. a degree of association above random expectation with their counterparts of a similar age. Black lines represent the minimum, median and maximum of the distribution. The blue and black horizontal lines represent the $25^{th}$, $50^{th}$ and $75^{th}$ percentiles. \label{fig_rgc_violin}}
\end{figure*}

\begin{figure*}
\includegraphics[width=0.49\textwidth]{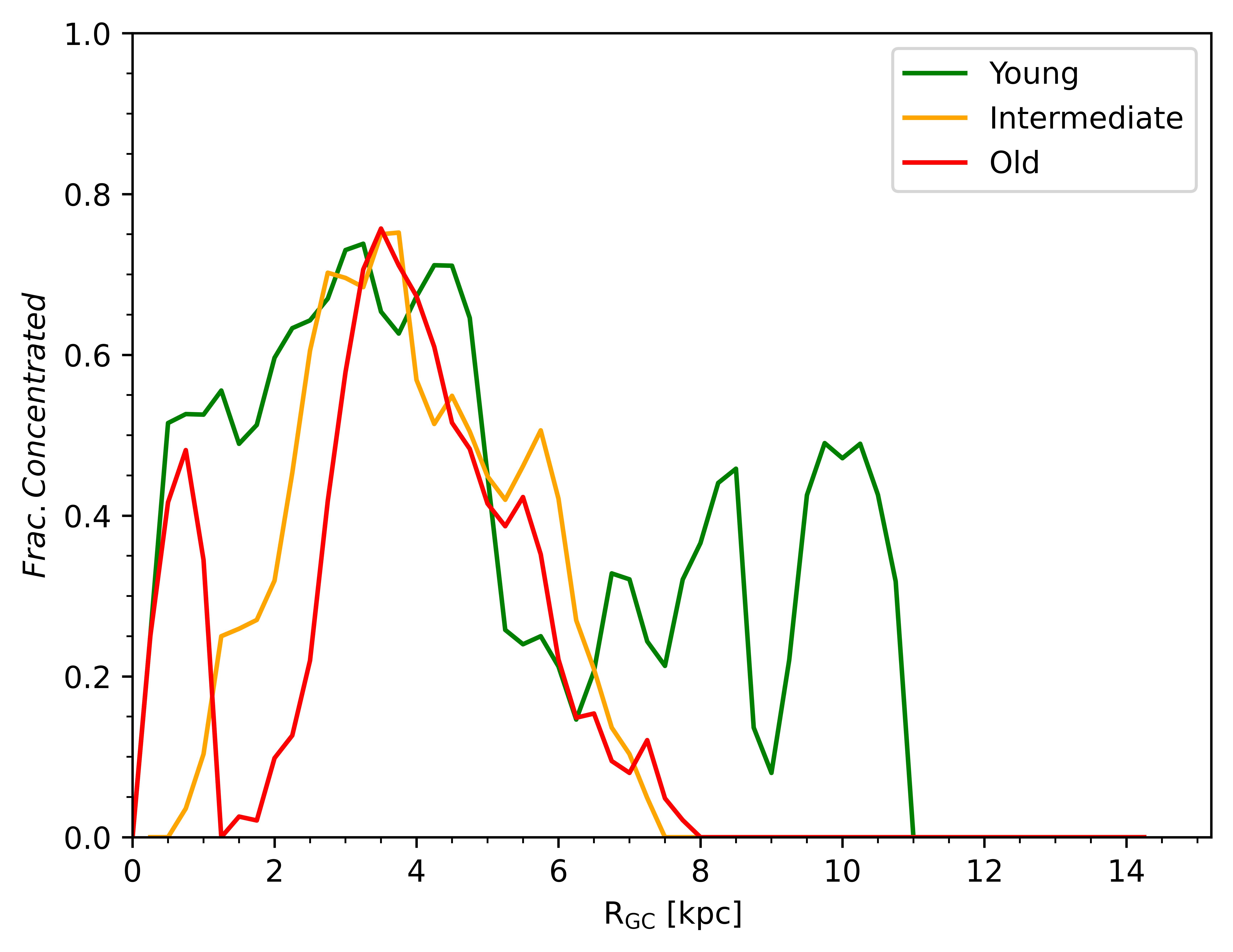}
\includegraphics[width=0.49\textwidth]{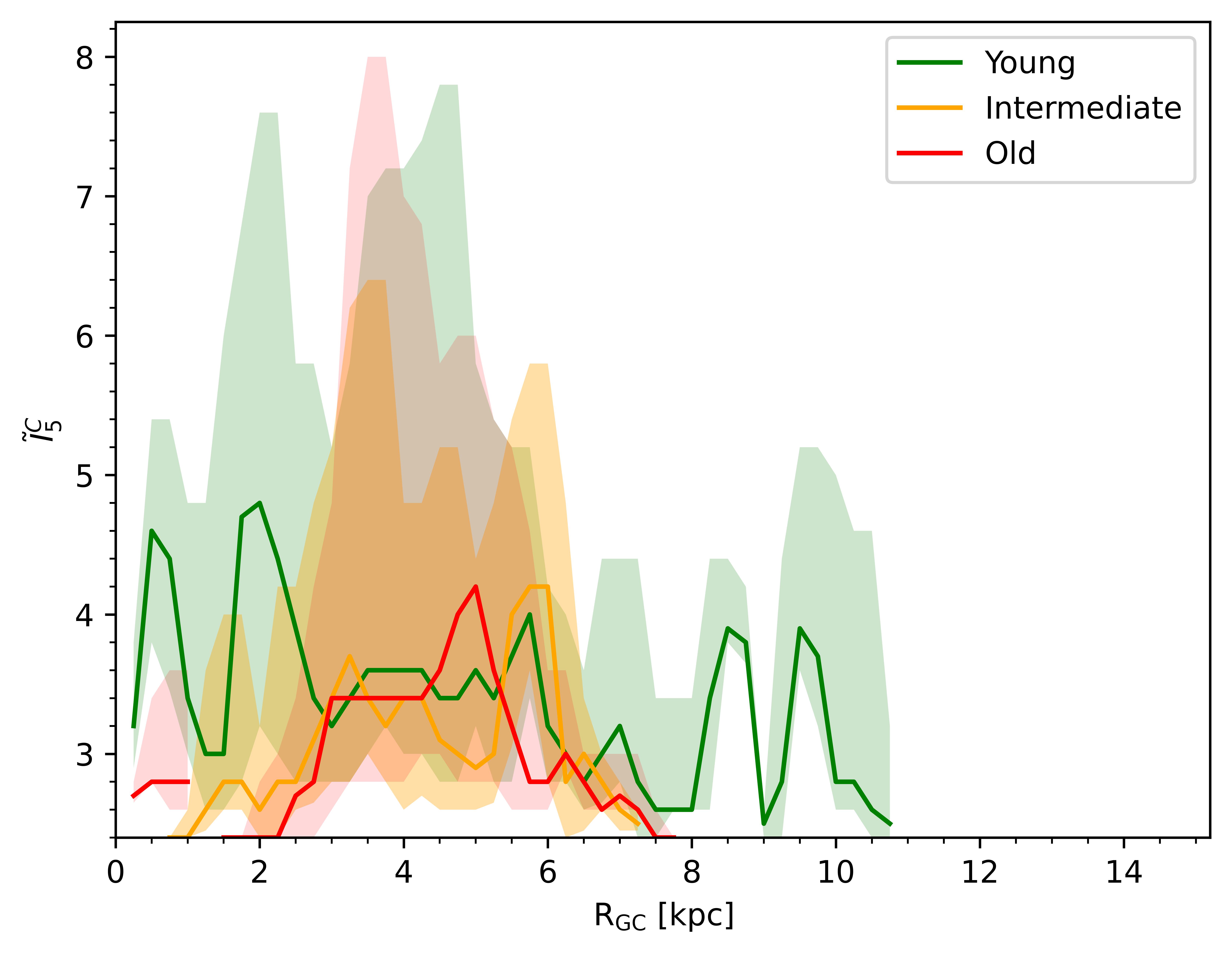}
\caption{Spatial clustering behaviour of Young, Intermediate and Old clusters as a function of galactocentric distance. The left panel shows the fraction of clusters found to be spatially concentrated with their counterparts of a similar age as discussed in Sect.\,\ref{sect_results_evo}; and the right panel shows the running 25th, 50th and 100th percentile of the $I_5$ index values of the spatially concentrated clusters.  \label{fig_rgc_i5}}
\end{figure*}

\subsubsection{Galactocentric distribution}\label{sect_results_evo_Rgal}

Due to the aforementioned variations across the disc, we examined how the spatial behaviour of the emerged clusters varies with galactocentric distance, R$_{\rm{GC}}$. We omit eYSCs from this part of the analysis as the difference in the shapes of the JWST/HST coverage area means we cannot be certain that any differences identified are attributable to real physical effects (i.e. evolutionary) or variations in the properties of regions that are not in both coverage areas (i.e. R$_{\rm{GC}}$ does not necessarily correspond to the same regions). 

Figure\,\ref{fig_rgc_violin} shows the frequency distribution of clusters in each group as a function of galactocentric distance for (i) all and (ii) concentrated clusters. For the former, there is no discernible difference in the radial distribution of clusters across the age groups. For the concentrated clusters, the median radial distance remains constant but the spread decreases with increasing cluster age narrowing to predominantly the R$_{\rm{GC}}\sim 2-6$kpc range. Figure\,\ref{fig_rgc_i5} shows the variation in the proportion of clusters spatially concentrated and their index values, for each age group as a function of galactocentric distance. Between R$_{\rm{GC}}\sim 2-6$\,kpc a similar proportion and concentration strength is observed irrespective of cluster age. However, outside this narrow range, variation is found with Young clusters exhibiting the greatest concentration behaviours.

The mechanism causing the lack of concentrated Intermediate/Old clusters at \(R_{\mathrm{GC}} > 6\)\,kpc is unclear. One possibility is that conditions in the outer disc favour the dissolution or decorrelation of spatial `structure', perhaps in a similar process that leads to the disintegration of strong two-arm spiral structure often seen outside co-rotation (\citealt{1992ApJS...79...37E}, \citealt{1995ApJ...445..591E}). However this would contrast with \citet{2024MNRAS.532.4583T}, who found clusters inside the co-rotation radius (\(R_{\mathrm{GC}} = 6.3\)\,kpc; \citealt{2021MNRAS.508..912S}) to be more easily disrupted. It may instead be that young clusters in the outer galaxy arise from isolated, localised star-formation events, such that their initial groupings disperse rapidly as they move away from their natal regions. Alternatively, this pattern could reflect gradual inside-out growth: over the past \(\sim10{-}100\)\,Myr the outer disc may have only recently achieved the gas surface densities necessary to form the observed structure. In this scenario, older tight concentrations were simply never formed here, whereas the inner disc ($<6.3$\,kpc) has sustained star formation over many tens to hundreds of Myr, producing the Intermediate and Old cluster concentrations now truncated near the co-rotation radius. A similar inside–out trend was reported by \citet{2010ApJ...712..858G} for NGC\,300: using resolved-star formation histories from colour–magnitude diagrams and $N$-body simulations, they found an increasing fraction of young stars with radius and negligible impact from radial migration, thereby favouring a growth‐driven rather than scattering‐driven scenario. The similarity in both the proportion of clusters that are concentrated and the strength of those concentrations across the age groups between $2$–$6$~kpc provides further evidence supporting a growth‐driven origin for NGC\,628.

For R$_{\rm{GC}}<2$\,kpc we find a more complex set of behaviours. For instance, from Fig.\,\ref{fig_rgc_i5}, we can see that while the fraction of concentrated Young clusters remains relatively high (at $\sim50\%$), there is a steady decrease of the fraction of concentrated Intermediate clusters towards smaller R$_{\rm{GC}}$. The Old clusters show yet a different behaviour, with a reasonably high fraction of concentrated clusters for R$_{\rm{GC}} < 1$\,kpc, but a clear dip between $\sim1-2$\,kpc. We verified that the double-peaked radial profile of the old cluster population persists when restricting the sample to clusters aged between 100~Myr and 1~Gyr. This check was performed to test whether the inner peak could be driven by contamination from globular clusters (ages~$\gtrsim$~1~Gyr). Although removing these objects markedly lowers the amplitude of the inner peak, the bimodal structure remains, confirming that the trend is intrinsic to the population of clusters younger than 1~Gyr. 

The intensity of the spatial concentrations also decreases substantially for the older age groups, contrary to that found generally across the disc (Sect.\,\ref{sect_results_evo}). More specifically, the Old and Intermediate population, when concentrated, are only very mildly so, with $I_5$ indices very close to the threshold, thus reducing the significance of the large fraction of Old concentrated clusters at R$_{\rm{GC}} \sim 1$\,kpc. For the Young clusters, however, there are significant peaks in the concentration indices for R$_{\rm{GC}}\sim0.5-1$\,kpc and R$_{\rm{GC}}\sim2$\,kpc. 

Our interpretation for these observed trends, lies in a comparison to some of the known dynamical features of this inner region. In particular, as shown in Fig.\,\ref{fig_RaDec_I5_age}, a radius of $\sim1.25$\,kpc corresponds to the transition between the galactic center and the beginning of the spiral arms. Just below this radii there is an inner Lindblad resonance, thought to be linked to the origin of the nuclear ring at $\sim$0.5-1\,kpc. The location of the first peak of concentrated Young clusters at $\sim$0.5-1\,kpc, would thus suggest this population of Young concentrated clusters being formed in the nuclear ring, but that the highly dynamical galactic centre quickly disperses these clusters as they evolve. Just outside of the inner Lindblad resonance (R$_{\rm{GC}}\simeq1.25$\,kpc) is an area which has been previously observed to have no H{\sc ii} regions, with a velocity drop around the galaxy's nuclear ring and inflowing (colder) gas \citep{2007A&A...466..905F}. Thus, it is reasonable to expect very little cluster structure to be formed here, and this is consistent with the dip in the concentration indices of the Young population at these radii. Moving outwards, spiral arms (which are regions of reduced shear) start to provide again the conditions to form tightly concentrated clusters, and we see this effect with a higher concentration index for Young clusters formed in the tips of the spiral arms. However, perhaps due to the harsh environment outside the spiral arm tips in this region (e.g. from strong shear and strong inward motions from the Lindblad resonance), any cluster structure would be prone to a very rapid dissolution as it emerges from the arms, so the overall proportion (and concentration index) of the Intermediate and Old cluster population at $<2$\,kpc radius remains lower than across the galactic disc. 



\subsection{Mass}\label{sect_results_mass}

\begin{figure*}
\includegraphics[width=0.49\textwidth]{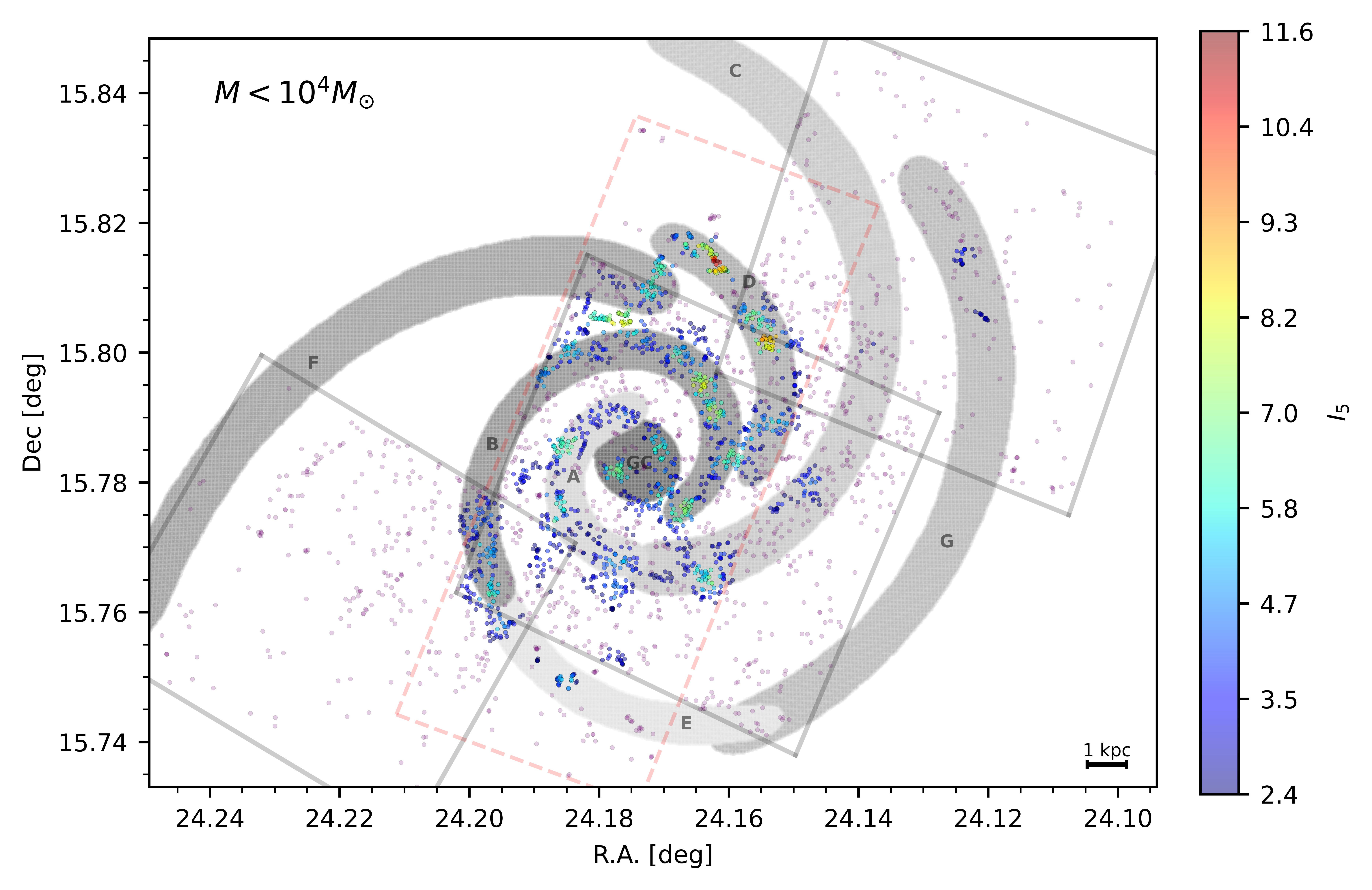}
\includegraphics[width=0.49\textwidth]{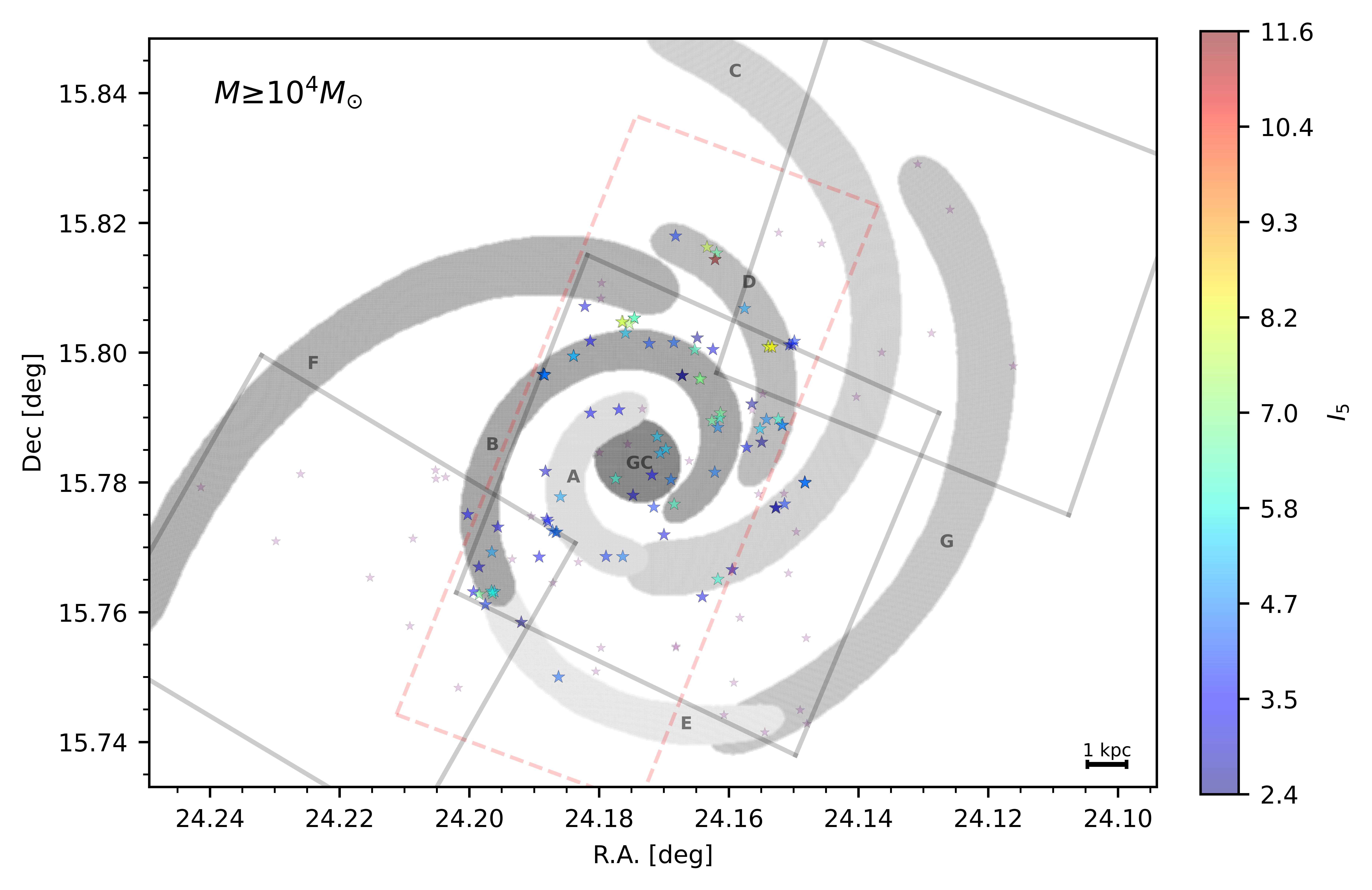}
\caption{Plots of index values, $I_5$, calculated by INDICATE for (Left:) Lower mass and (Right): YMCs, to investigate if Type 2 mass segregation is present as described in Sect.\,\ref{sect_results_mass}. As in previous figures, the grey and red-dashed panels denote the position of the HST and JWST images respectively. Clusters with an index value above the significance threshold ($I_5 > 2.4$) are colour-coded by their $I_5$ index, where higher values denote greater degrees of spatial concentration. Purple markers are clusters with an index value below the significance threshold. \label{fig_HMLM_I5_T2}}
\end{figure*}

To quantify how the clusters' spatial behaviour might change as a function of their mass, we compare the distribution of clusters with high ($\ge10^{4}\,M_{\odot}$) and lower ($<10^{4}\,M_{\odot}$).  Two quite different realisations of mass segregation are explored here. The classic ‘Type 1’ definition examines if massive clusters are spatially concentrated with other massive clusters, significantly more so than seen for their lower mass counterparts. Another definition, ‘Type 2’, looks at whether massive clusters are in concentrated regions, but are not necessarily concentrated with other massive clusters. Due to the age-mass bias (Sect.\,\ref{sect_bias}) it is incorrect to simply divide the catalogue into ‘high’ and ‘lower’ mass, as lower mass clusters will inherently represent a younger population and high mass clusters an older population. Instead, we apply an age-mass cut to the catalogue to remove this bias. The new sample consists of clusters with an age of $\le100$\,Myr and a mass larger than the minimum mass of clusters $>100$\,Myr ($\sim790\,\text{M}_{\odot}$), visualised in the right panel of Fig.\,\ref{fig_cat_agemass_cuts}. A total of 3087 clusters meet this criteria, of which 137 are high mass (thus corresponding to YMCs).  

To determine if Type 1 mass segregation is present, we divide the sample into two sub-samples (‘high’ and ‘lower’ mass), and run INDICATE on both separately. If mass segregation is present, we would expect the high mass sample (YMCs) to exhibit significantly greater tendencies to spatially associate than the lower mass sample. No signatures were found: YMCs do not have a tendency to concentrate together in NGC\,628. In fact, the $I_5$ values suggest the opposite - massive clusters tend to be more spatially dispersed w.r.t. each other than lower mass clusters. A 2 sample K-S test confirms that while Type 1 mass segregation is not present, YMCs and lower mass clusters do have different spatial behaviour ($p<0.05$). Moreover, we then examined the behaviour of the Old high mass clusters ($>100$\,Myr) finding that while there is no significant change in the tightness of the concentrations, the proportion of the population that are concentrated decreases by $17.5\%$.  

To determine if Type 2 mass segregation is present, INDICATE is run on the full sample (within our age-mass cut). Again, if present, YMCs should exhibit significantly greater tendencies to spatially associate than the lower mass clusters. We find that a greater proportion of massive clusters ($67.2\%$) are found in concentrated regions compared to lower mass clusters ($57.7\%$). As shown in Fig.\,\ref{fig_HMLM_I5_T2}, the degree of association is typically higher for YMCs compared to their lower mass counterparts (2-3 clusters more in their local neighbourhood) because they tend to be located in regions where the lower mass population exhibits greatest spatial association. Moreover, massive clusters that have signatures of Type 2 mass segregation are almost exclusively located within the spiral arms and galactic center - whereas those in the interarm regions typically do not. This result is suggests most of the massive clusters are forming in the high-density, shock-compressed spiral arms, in regions of intense star formation activity. Occasionally, an exceptionally massive GMC in the inter-arm region will collapse and form a massive cluster but,as overall star formation intensity is low there, the cluster has very few neighbors nearby and relatively isolated (no Type 2 signal). A 2 sample K-S test comparing the index values of the two populations confirms the found signatures of Type 2 mass segregation are real ($p<0.05$).


\section{Conclusions}\label{sect_conclude}

We have characterised the spatial distributions of star clusters in the nearby face-on spiral galaxy NGC\,628. The aim of our study was to quantify the impact galactic environment has on the formation, early and longer-term evolution of stellar clusters, and YMCs in particular. This was achieved through the analysis of a complete sample, from FEAST collaboration, of the cluster population consisting of 6890 candidates from the (partially) embedded phase to $>100\,\text{Myr}$.  

The powerful novel statistical clustering tool INDICATE by \citet{2019A&A...622A.184B} was used to characterise the spatial behaviour of the clusters. INDICATE is a local indicator of spatial association, meaning that it quantifies the clustering tendencies of each cluster individually. As such, it is possible to not only study general variations with cluster properties (age, mass) in the galaxy, but also to compare the impact disc position has between clusters with similar ages and masses. 

We found that the spatial behaviour of clusters in NGC\,628 varies with galactic position, evolutionary stage and mass. The majority of young clusters $\le10$Myr, which are still embedded within their natal clouds, are spatially concentrated. Typically they tend to be in the tightest spatial associations, located in multiple ‘hot spots’ across the spiral arms, with the most intense concentrations occurring in arm D. These locations correlate well with regions that have the highest molecular star formation efficiency\footnote{defined as the ratio of the star formation rate from H$\alpha$ emission to molecular gas mass surface density, i.e. $\sum_{\rm{ SFR(H\alpha)}}/\sum_{mol}$} as found by \citet{2024MNRAS.531..815L}, suggesting multiple clusters are forming in these tight "structures" and that the spatial behaviour we identify is a star formation signature rather than indicative of early dynamical evolution.  Once emerged, clusters of a similar age do still tend to concentrate together within the spiral arms, but in looser spatial associations than observed for the embedded population. We speculate that this could be due to clusters moving away from their birth place as their natal cloud is cleared, and/or due to the fact that some clusters may become dispersed once they loose the gravitational potential from the gas of their natal cloud. While the median degree of association remains approximately constant with increasing cluster age after the clusters are fully emerged, the proportion of clusters which are spatially concentrated decreases, reaching a minimum at ages $>100$\,Myr. This result is in good agreement with \citet{2015ApJ...815...93G} who, by applying the 2-point correlation function to an earlier (and less complete) version of our cluster sample, found that the strength of the clustering decreases as age increases and that clusters $>40$\,Myr tend to be randomly distributed.  

The radial distribution of all emerged clusters across the disc is similar irrespective of cluster age, but that of the concentrated population progressively narrows to within a radius of $2-6$\,kpc of the galactic center with increasing age. Within this radius a similar proportion of clusters that are concentrated, and the strength of this concentration, is found across all age groups. This suggests either tight cluster spatial `structure' in the outer disk (1) is more easily disrupted than the inner disk (not supported by previous studies); (2) is formed in isolated, localised star-formation events, such that their initial groupings disperse rapidly; (3) is a result of a gradual inside-out growth scenario for NGC\,628, such that the disc outside the galactic co-rotation radius ($>6.3$\,kpc) has built up over the past $\sim10-100$Myr to gas surface densities capable of forming these structures, leading to the observed truncation beyond the co-rotation radius. We also found noticeable changes in the spatial behaviour of clusters of different ages in the galactic centre, for radii $<2$\,kpc, which we interpret as being related to the location of an inner Lindblad resonance (at $\lesssim $1\,kpc), the nuclear ring (at $\sim$0.5-1\,kpc), and the start of the spiral arms (at $\sim$1.25-2\,kpc), and suggest that this area's shear and particularly strong radial motions, make it hard not only for clusters to form in tight concentrations but even more so in remaining tightly concentrated as they evolve. 

Finally, we looked for signatures of mass segregation in NGC\,628. Massive clusters ($>10^4\,\text{M}_{\odot}$) were found to typically be spatially dispersed w.r.t. each other, in agreement with \citet{2015ApJ...815...93G}, and this was most pronounced at ages $>$100\,Myr. By using INDICATE, we were able to extend this analysis to examine the spatial relationship between massive cluster and their lower mass counterparts. We found that YMCs in the spiral arm and galactic center regions tended to be associated with the most intense spatial concentrations of lower mass cluster population, whereas those in the inter-arm regions tended to be relatively isolated. 

Our results suggest a `big picture' scenario in NGC\,628 where the process of hierarchical fragmentation and collapse of GMCs tend to produce a higher mass cluster and multiple lower mass clusters in relatively close proximity in regions that have a high star formation activity, especially in the inner galaxy ($<6.3$\,kpc). In regions of lower star formation activity, massive clusters will form in relative isolation, potentially due to the collapse of a somewhat rare high-density cloud. Once clusters have fully emerged, this structure is mostly retained but diluted, as clusters are dynamical dispersed from their birth place and/or disrupted due to insufficient stellar mass or stellar depletion. This may reflect either faster dispersal of isolated tight cluster spatial ‘structure’ in a lower gas density outer disk or gradual inside-out growth, with the formation of this structure shifting outward over time. A comprehensive comparison between the spatial distribution of young embedded clusters and that of the molecular gas is required to better understand these observations, and will be presented in a forthcoming companion paper.


\section*{Acknowledgments}
We thank the anonymous referee for their insightful comments and helpful suggestions, which greatly improved this manuscript. ASMB and ADC acknowledge the support from the Royal Society University Research Fellowship URF/R1/191609 (PI: ADC). The calculations used in this work were performed using the supercomputing facilities at Cardiff University operated by Advanced Research Computing at Cardiff (ARCCA) on behalf of the Cardiff Supercomputing Facility and the HPC Wales and Supercomputing Wales (SCW) projects. This work is based in part on observations made with the NASA/ESA/CSA James Webb Space Telescope, which is operated by the Association of Universities for Research in Astronomy, Inc., under NASA contract NAS 5-03127. The data were obtained from the Mikulski Archive for Space Telescopes (MAST) at the Space Telescope Science Institute. These observations are associated with program~\#1783. Support for program~\#1783 was provided by NASA through a grant from the Space Telescope Science Institute, which is operated by the Association of Universities for Research in Astronomy, Inc., under NASA contract NAS 5-03127. The specific observations analyzed can be accessed via \href{http://dx.doi.org/10.17909/f4vm-c771}{doi:10.17909/f4vm-c771}.  
AA and AP acknowledge support from the Swedish National Space Agency (SNSA) through the grant 2021-00108. AA and HFV acknowledge support from SNSA 2023-00260. KG is supported by the Australian Research Council through the Discovery Early Career Researcher Award (DECRA) Fellowship (project number DE220100766) funded by the Australian Government.


\section*{Data Availability}
The INDICATE tool is available to download from GitHub at https://github.com/abuckner89/INDICATE. The data underlying this article will be shared on reasonable request to the corresponding author.


\bibliographystyle{mnras}
\bibliography{TSESG1} 


\newpage
\appendix

\section{Age-Mass Distribution}
\subsection{Impact of age-mass bias on derived statistics}\label{sect_app1}

\begin{figure*}
\includegraphics[width=0.45\textwidth]{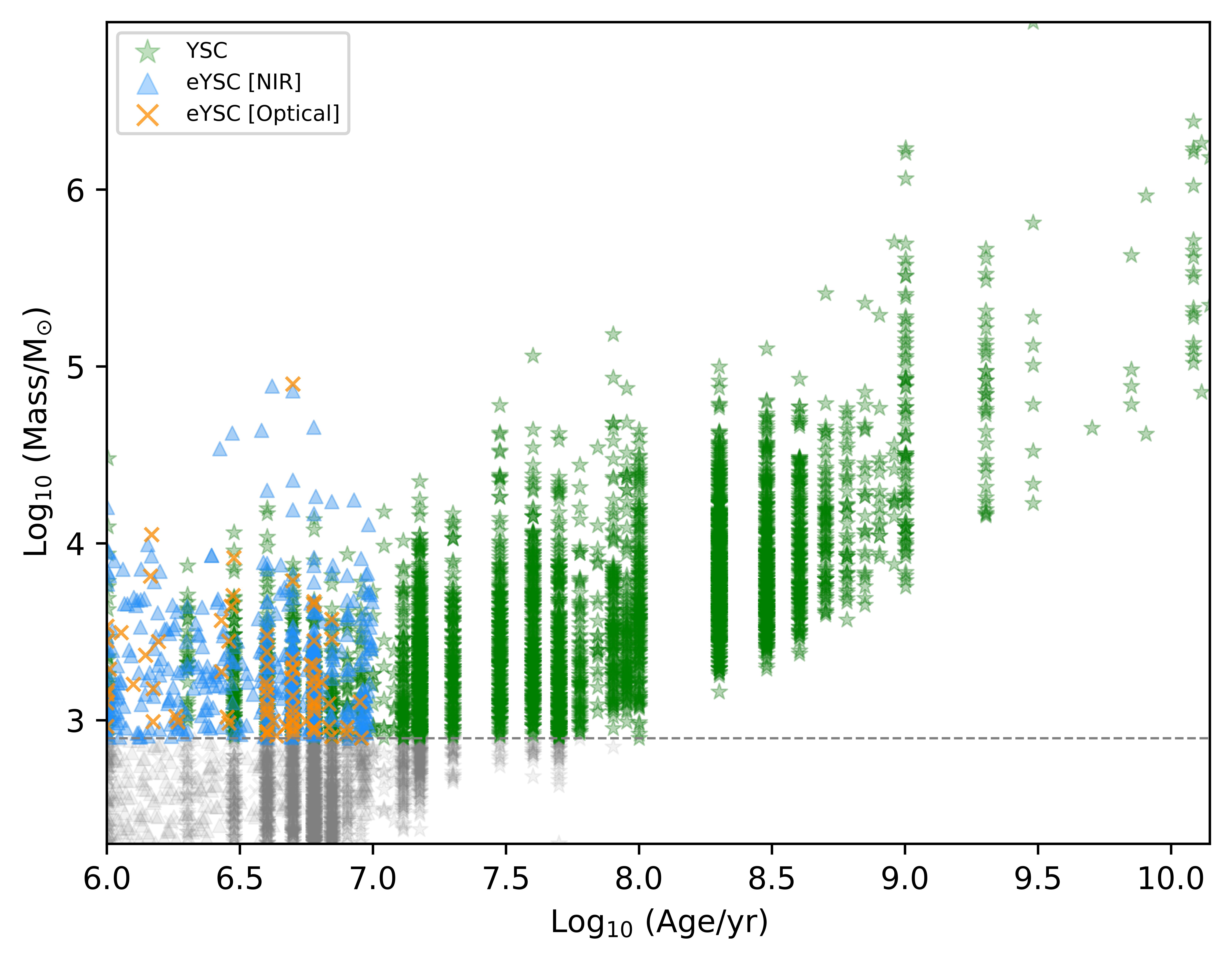}
\includegraphics[width=0.45\textwidth]{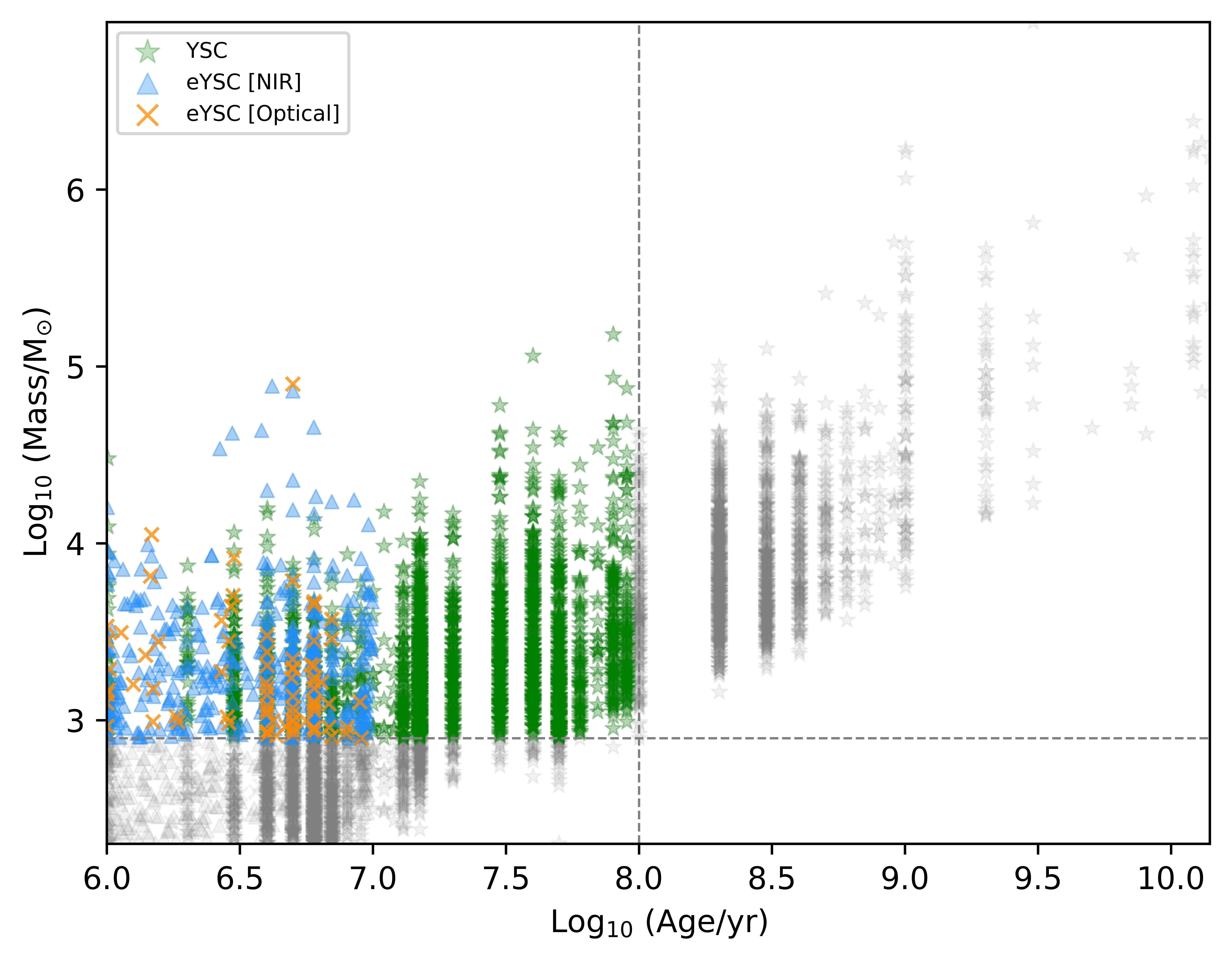}
\caption{Plot of age and mass distributions for our (Left:) mass-restricted and (Right:) age-restricted versions of our cluster catalogue as shown in Figs.\,\ref{fig_cat_agemass} and \ref{fig_cat_agemass_hists}. Grey markers in the top panels represent clusters which were excluded under the restriction criteria.}\label{fig_cat_agemass_cuts}
\end{figure*}

We tested the need to apply cuts to the cluster sample, owing to the Age-Mass bias present. INDICATE was run on three versions of the cluster sample:

\begin{enumerate}
    \item Full: no cuts were made.
    \item Mass restricted:  clusters with a mass less than the minimum mass of clusters that are $>100$\,Myr ($<790.1\,\text{M}_{\odot}$) were removed 
    \item Age restricted: an age limit of 100\,Myr was applied to the mass restricted sample 
\end{enumerate}

\begin{figure*}
\includegraphics[width=0.45\textwidth]{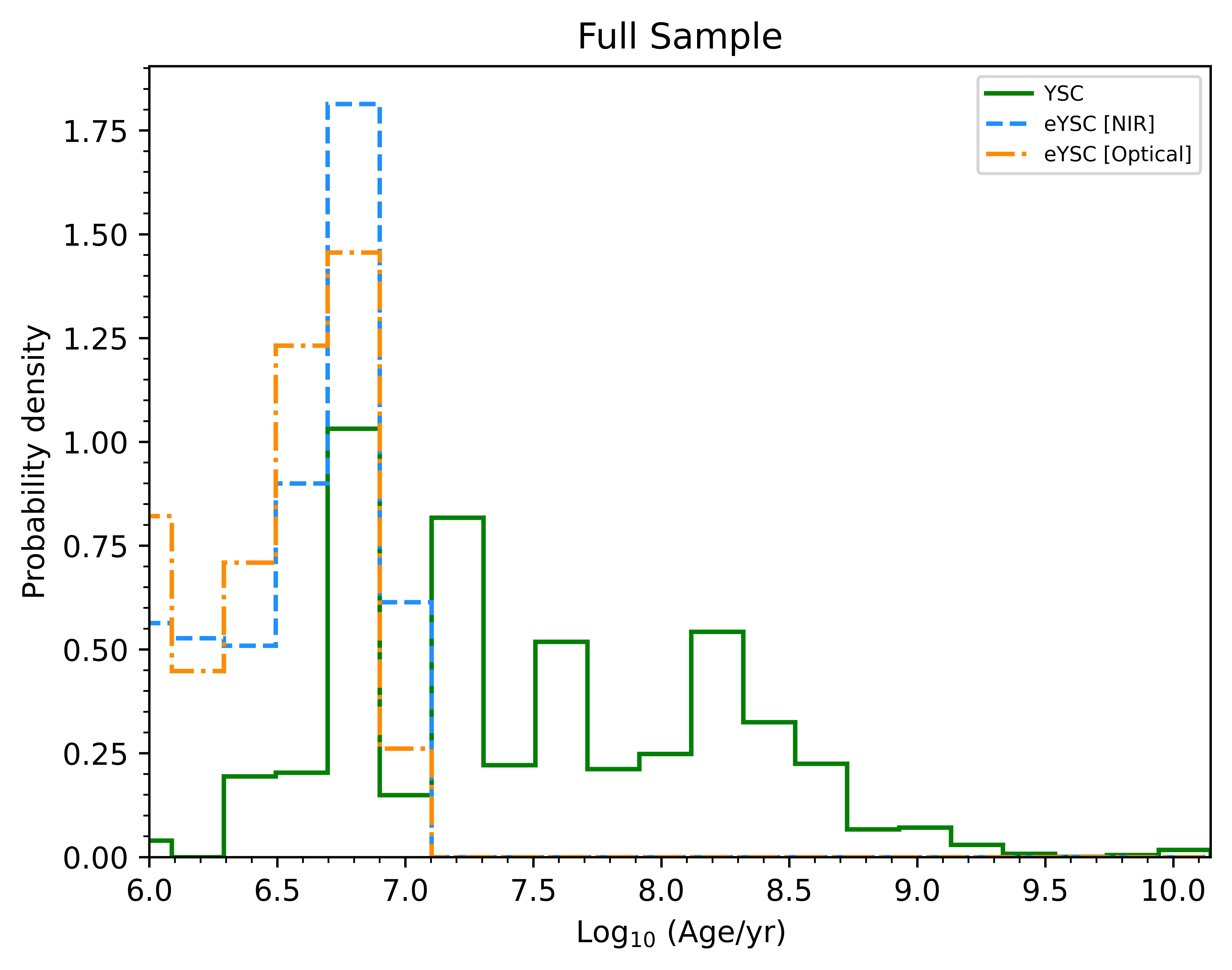}
\includegraphics[width=0.45\textwidth]{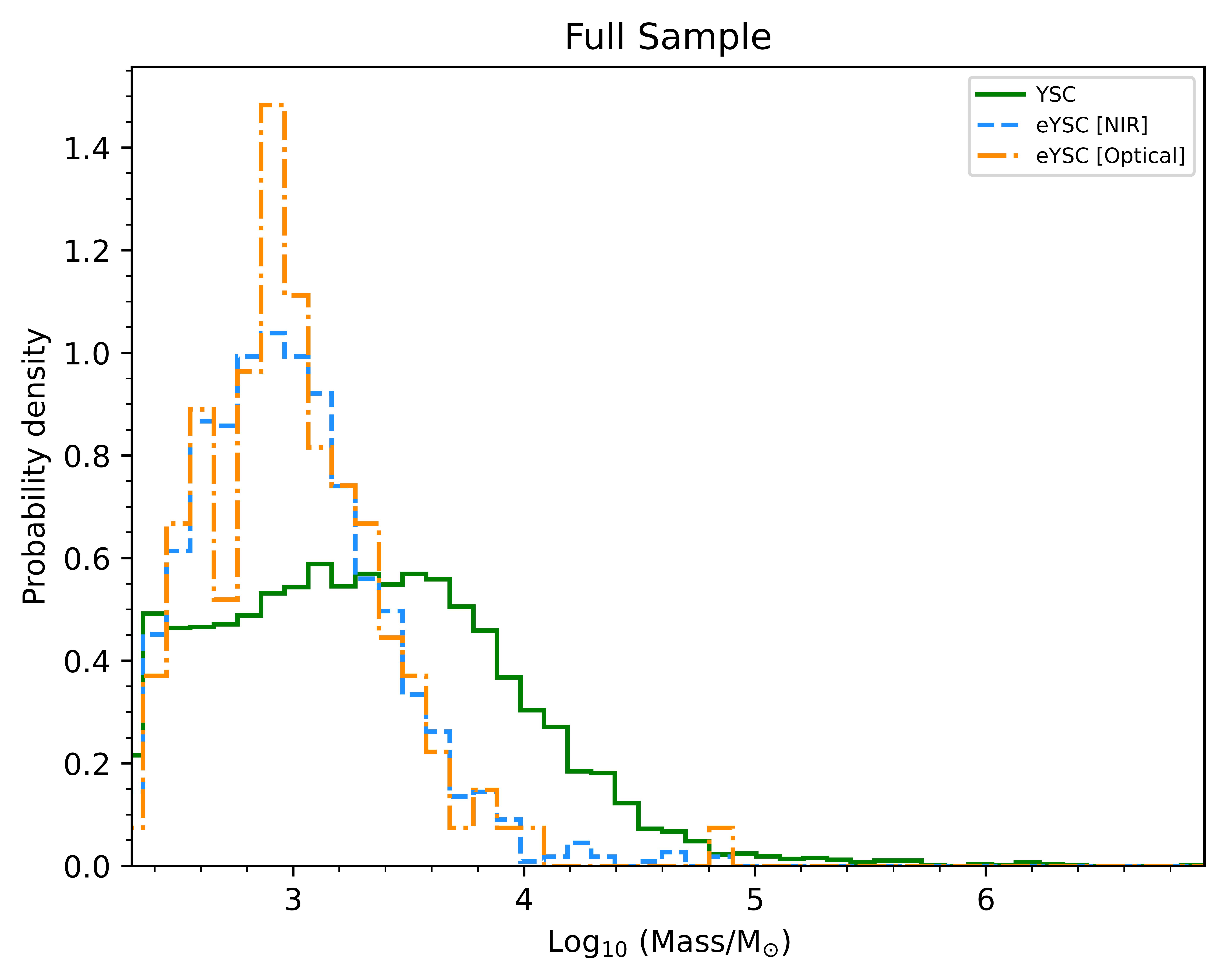}
\includegraphics[width=0.45\textwidth]{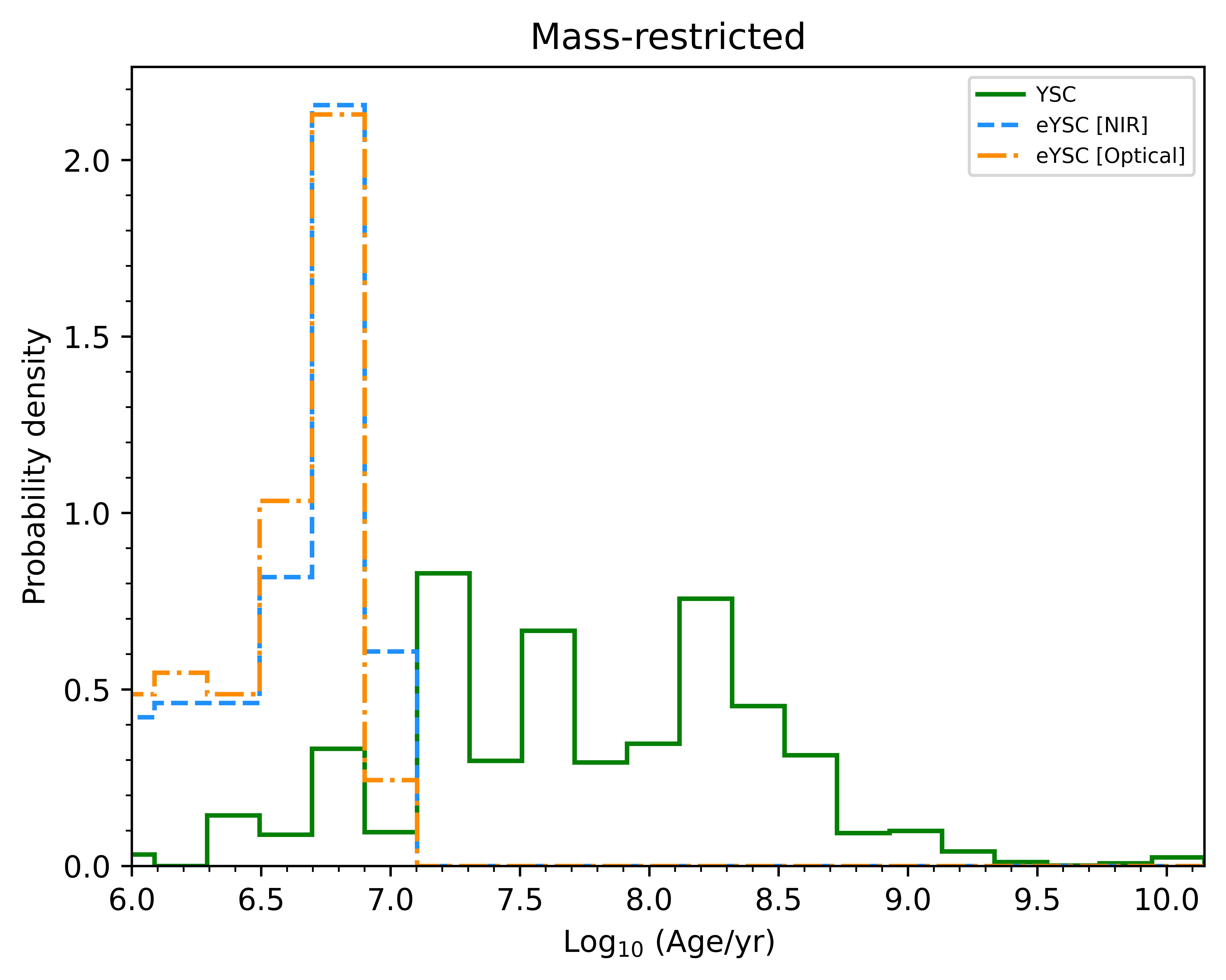}
\includegraphics[width=0.45\textwidth]{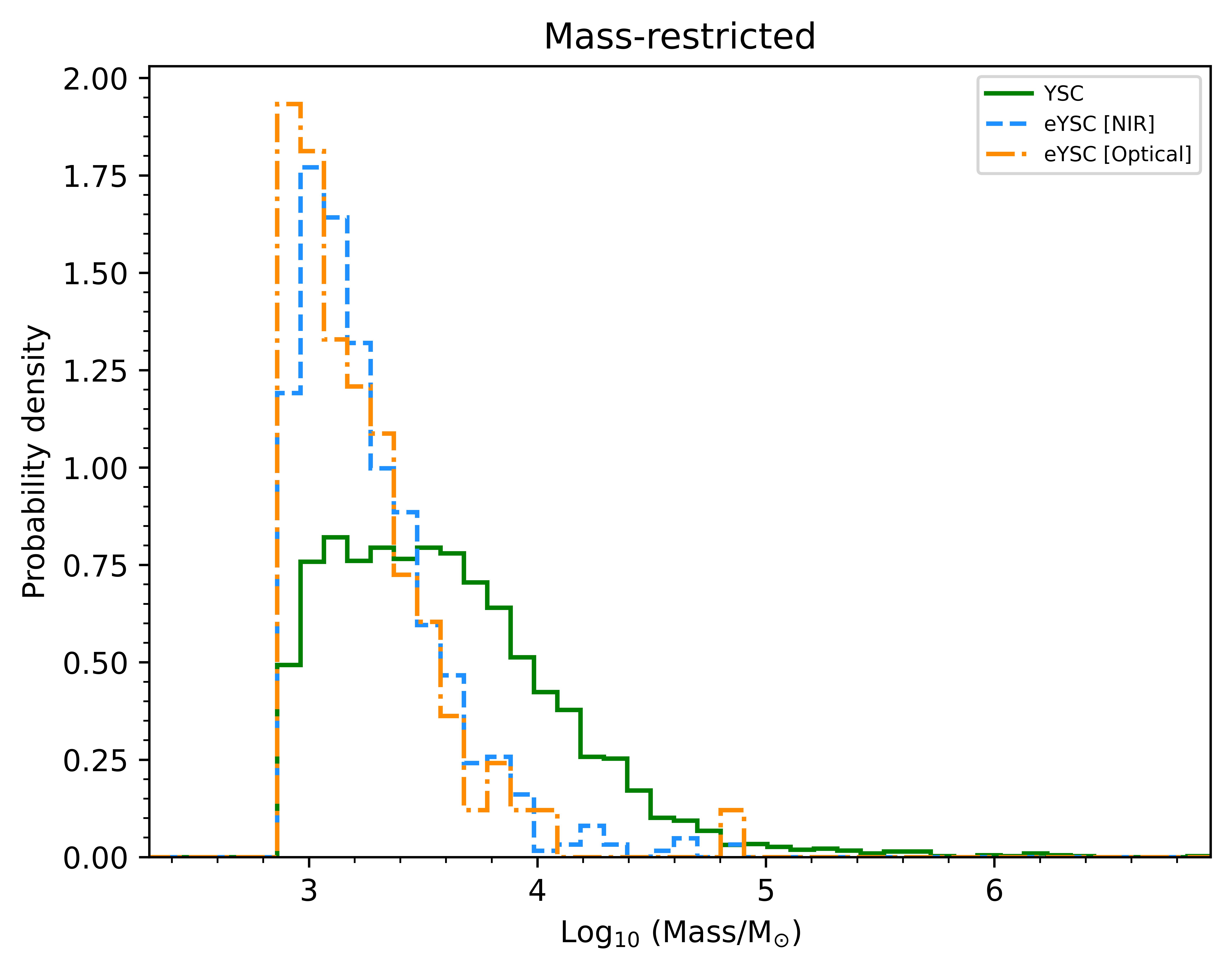}
\includegraphics[width=0.45\textwidth]{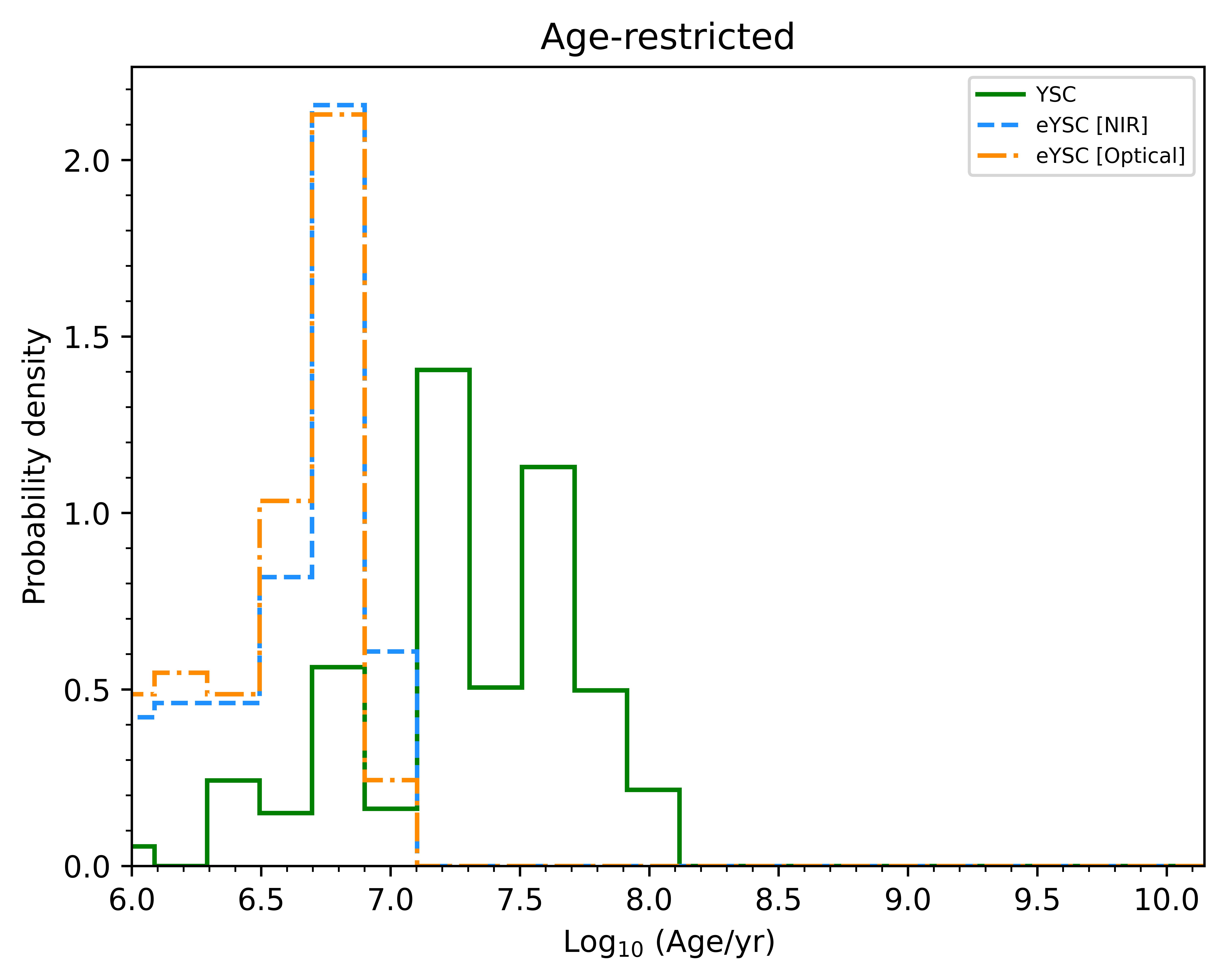}
\includegraphics[width=0.45\textwidth]{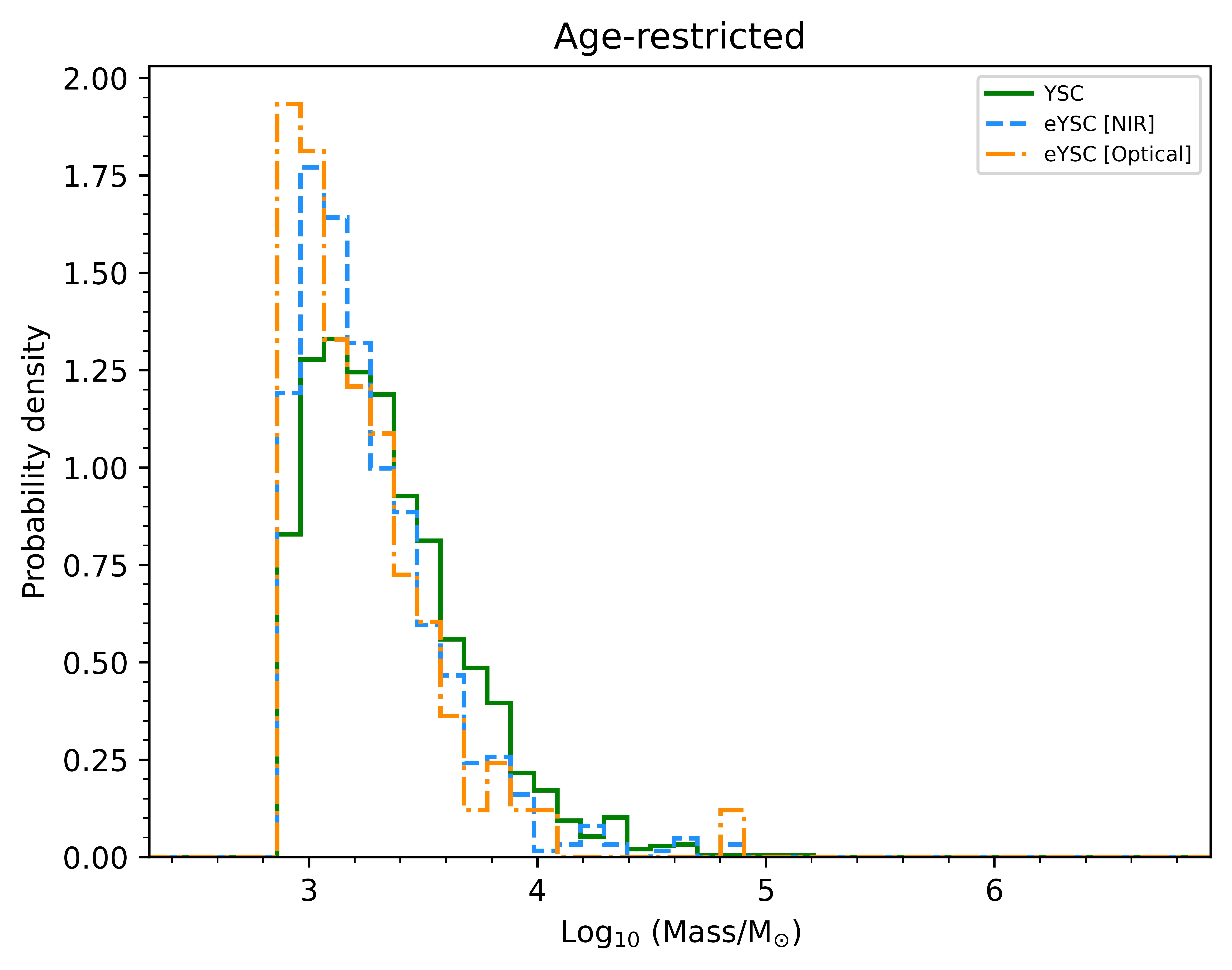}
\caption{Histograms for the full, mass-restricted and age-restricted cluster samples showing (left:) age and (right) mass distributions. The histograms have been normalised as probability densities (such that the total area under the bars of each histogram equals one) to enable direct comparison of the age and mass distributions of the eYSC-NIR, eYSC-Optical and YSC samples.\label{fig_cat_agemass_hists}}
\end{figure*}

Figures\,\ref{fig_cat_agemass_cuts} and \ref{fig_cat_agemass_hists} visualise the full and two restricted sample versions of the catalogue. Additionally we created mass-restricted versions of our Young and Intermediate age samples (described in Sect\,\ref{sect_results_evo}). Table\,\ref{tab_tests} gives statistics on the index values derived for each sample. 

In all samples, either no or small variations occurred in the median index values and/or $\%$ of spatially concentrated clusters. The conclusions drawn about the general spatial behaviour of the clusters are unchanged between the full and restricted versions of the sample. 

In cases where a variation did occur, the strength of the spatial behaviour signatures increased rather than decreased. The greatest variations occurred in the age-restricted and Young samples. This was expected, as the most tightly spatially concentrated clusters tend to be young and thus the restrictions inherently impact the youngest clusters the most (increasing larger proportions of population were removed with decreasing cluster age). For the age-restricted sample (versus the respective full sample) this meant that the dispersed older population was removed, so stats represent the spatial behaviour of a younger population which tend to have greater spatial associations than their older counterparts (Sect.\,\ref{sect_results_evo}). For the Young samples, index values from the mass-restricted samples represent the spatial behaviour of a higher-mass population, which tend to have greater spatial associations than their lower-mass counterparts (Sect.\,\ref{sect_results_mass}).

\begin{table}
\begin{center}
\caption{Table detailing the differences in the percentage of clusters found to be spatially concentrated and their median index value ($\tilde{I}^{C}_5$) for the full, mass-restricted and age-restricted samples. \label{tab_tests}}
\begin{tabular}{l l c c c} 

 \hline
Sample & Restriction & No. Clusters & $\%$ Concentrated & $\tilde{I}^{C}_5$ \\ 
 \hline\hline
\multirow{3}{*}{Full} & None & 6890 & 51.8 & 3.8 \\
& Mass-restricted & 4755 & 51.8 & 3.8 \\
& Age-restricted & 3087 & 58.1 & 4.2 \\
\hline
\multirow{2}{*}{Young} & None & 2805 & 49.4 & 3.4 \\
& Mass-restricted & 1255 & 55.7 & 3.8 \\
 \hline
\multirow{2}{*}{Intermediate} & None & 1201 & 41.2 & 3.2 \\
& Mass-restricted & 1143 & 44.3 & 3.2\\
 \hline
\end{tabular}
\end{center}
\end{table}

\subsection{Sub-sample statistics}\label{sect_app2}
\begin{figure*}
\includegraphics[width=0.45\textwidth]{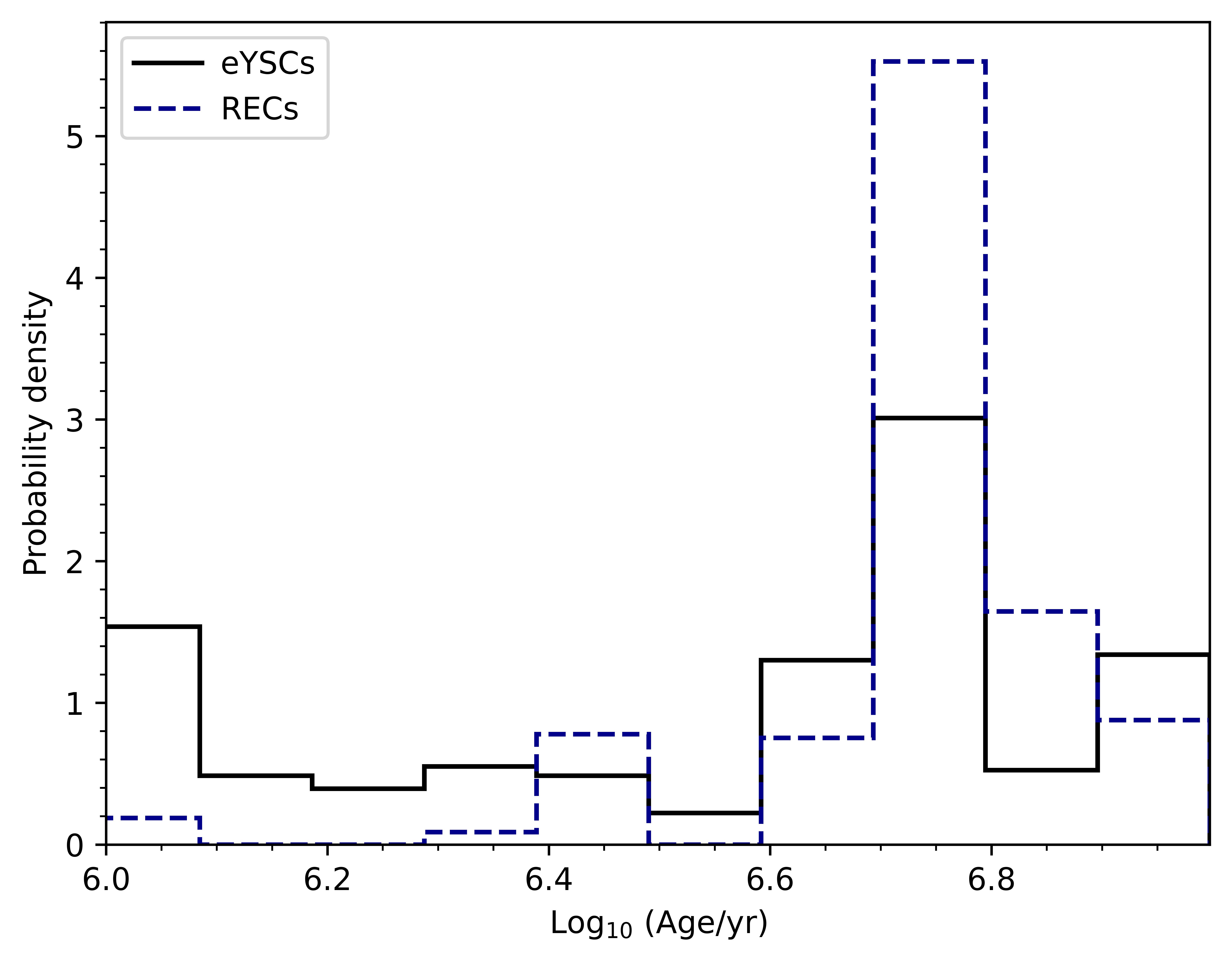}
\includegraphics[width=0.45\textwidth]{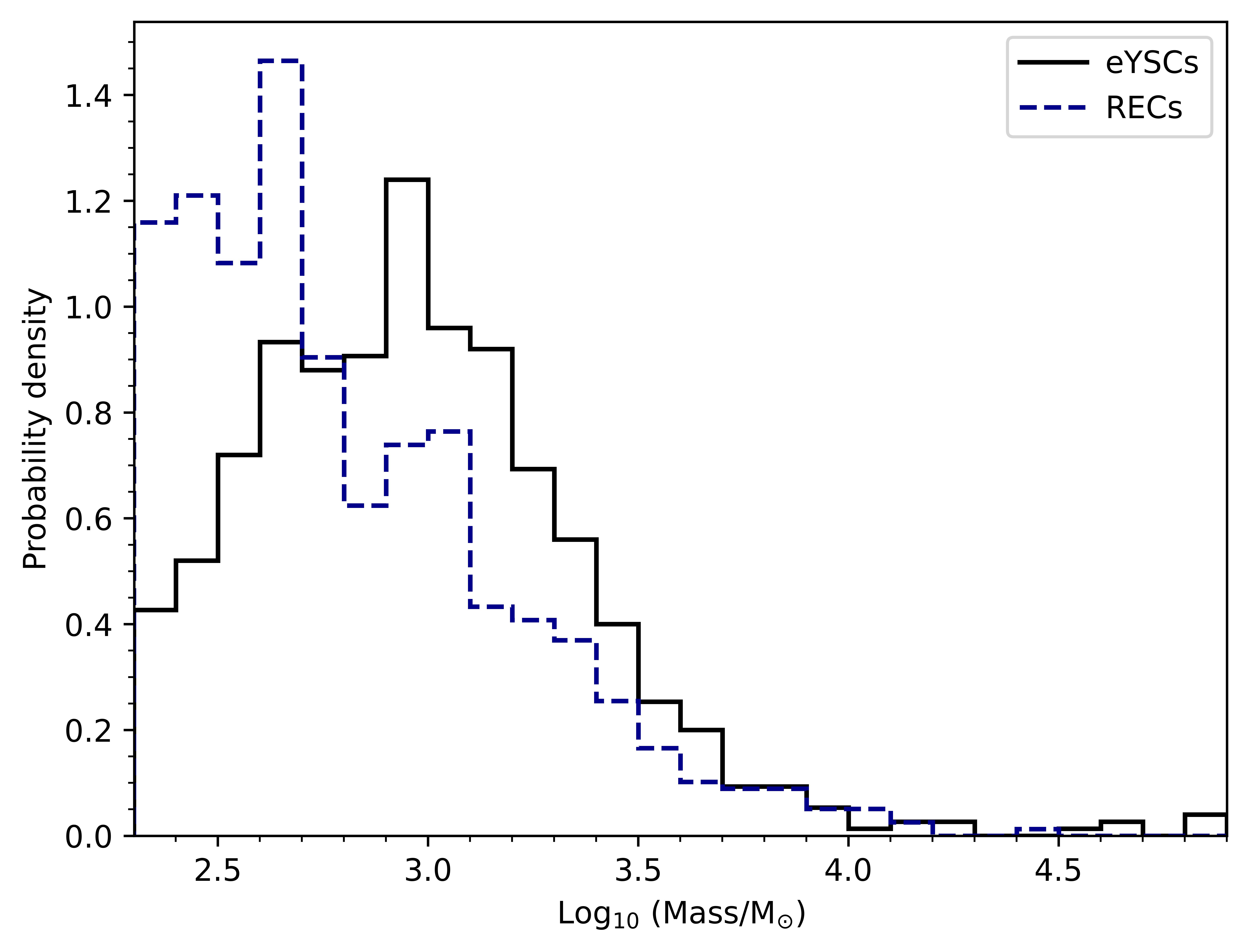}
\caption{Probability–density histograms comparing the eYSCs and RECs: (left) $\log_{10}(\mathrm{age/yr})$ and (right) $\log_{10}(M/M_\odot)$. Both panels are normalised to unit area and use identical binning to enable direct comparison; the $y$–axis shows probability density. See Section~\ref{sect_results_eYSCs}.\label{fig_hist_embedded}}
\end{figure*}

\begin{figure*}
\includegraphics[width=0.45\textwidth]{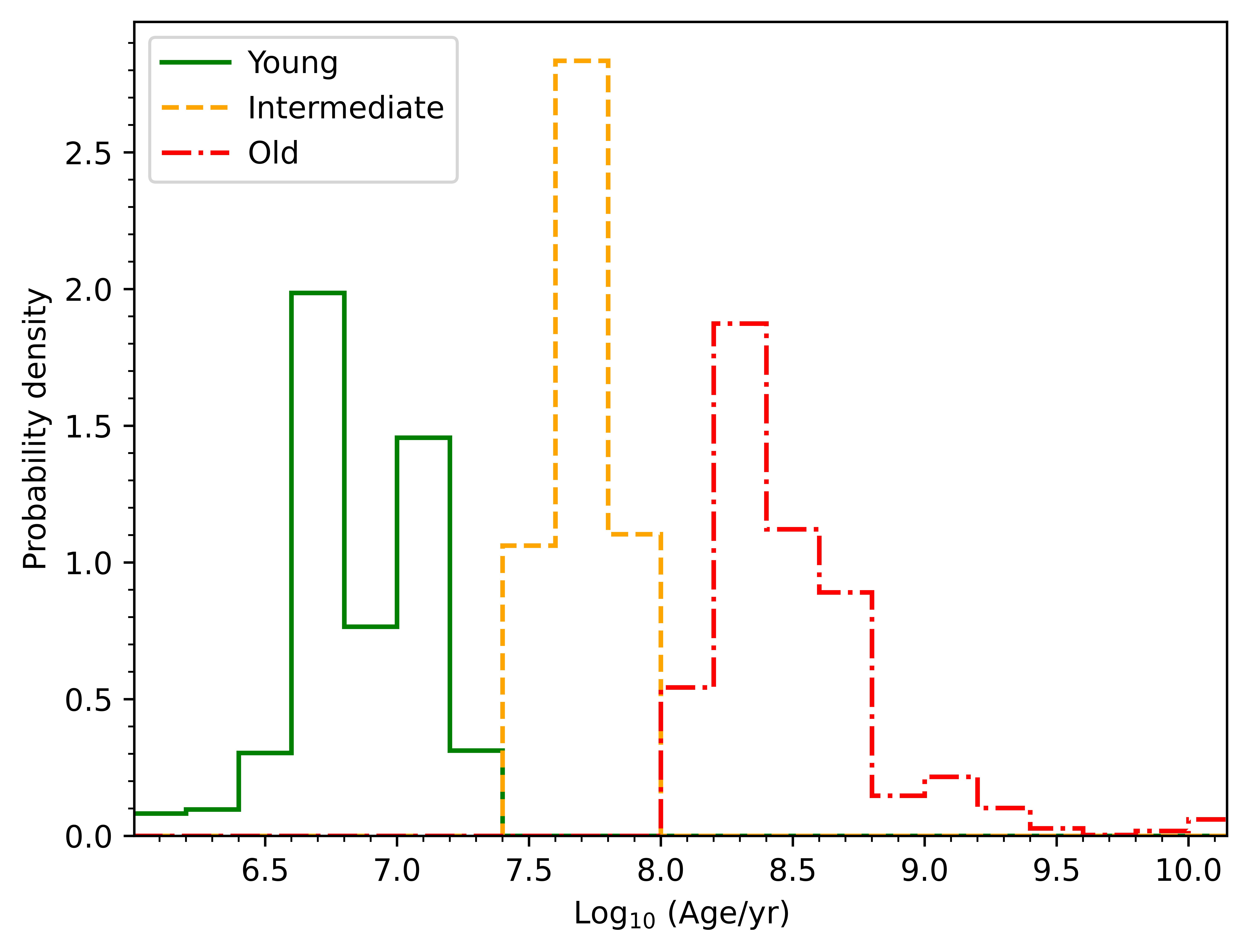}
\includegraphics[width=0.45\textwidth]{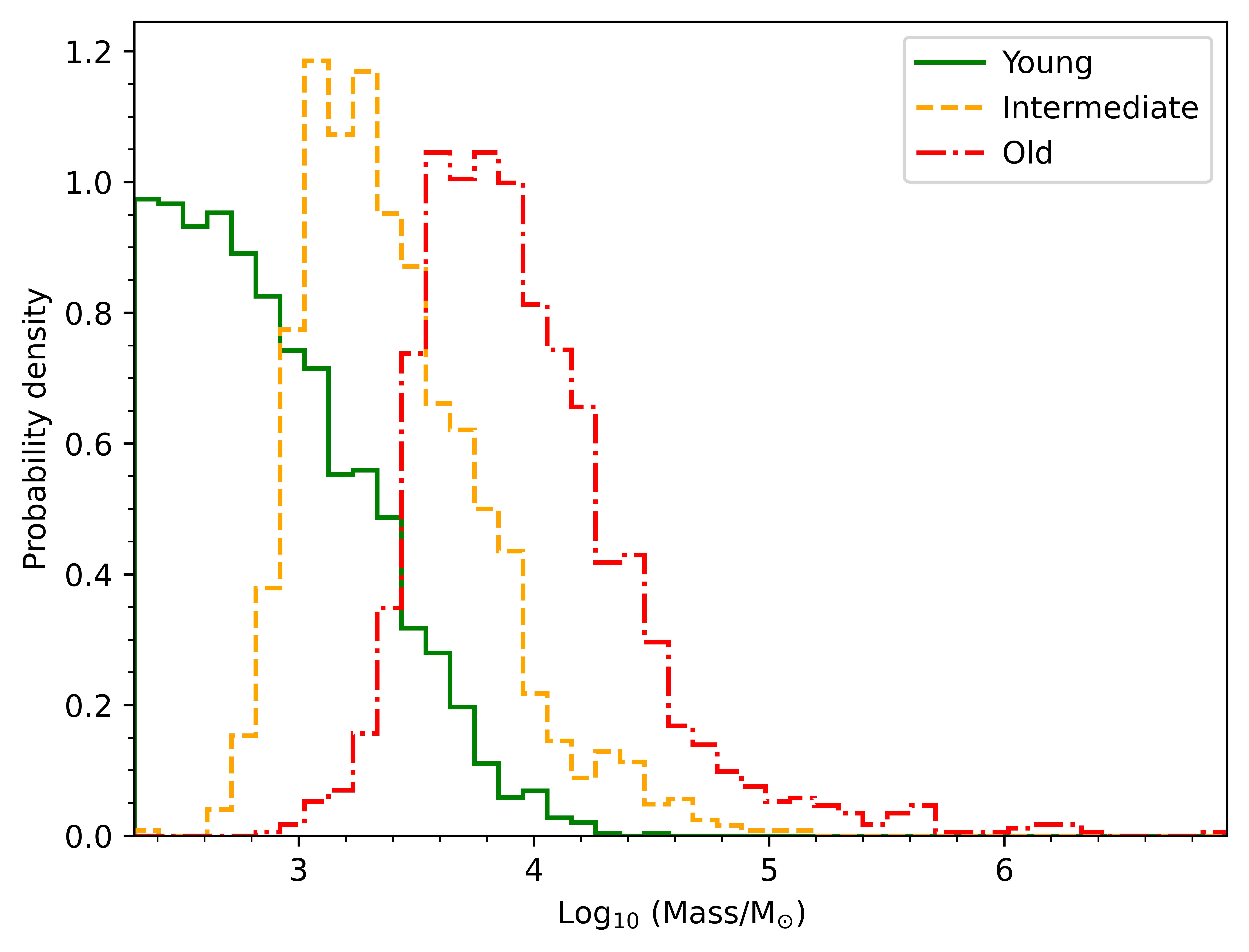}
\caption{Probability–density histograms for the evolutionary subsamples (Young, Intermediate, Old): (left) $\log_{10}(\mathrm{age/yr})$ and (right) $\log_{10}(M/M_\odot)$. Both panels are normalised to unit area with identical binning; the $y$–axis shows probability density. See Section~\ref{sect_results_evo}.\label{fig_hist_ages}}
\end{figure*}

\begin{figure*}
\includegraphics[width=0.45\textwidth]{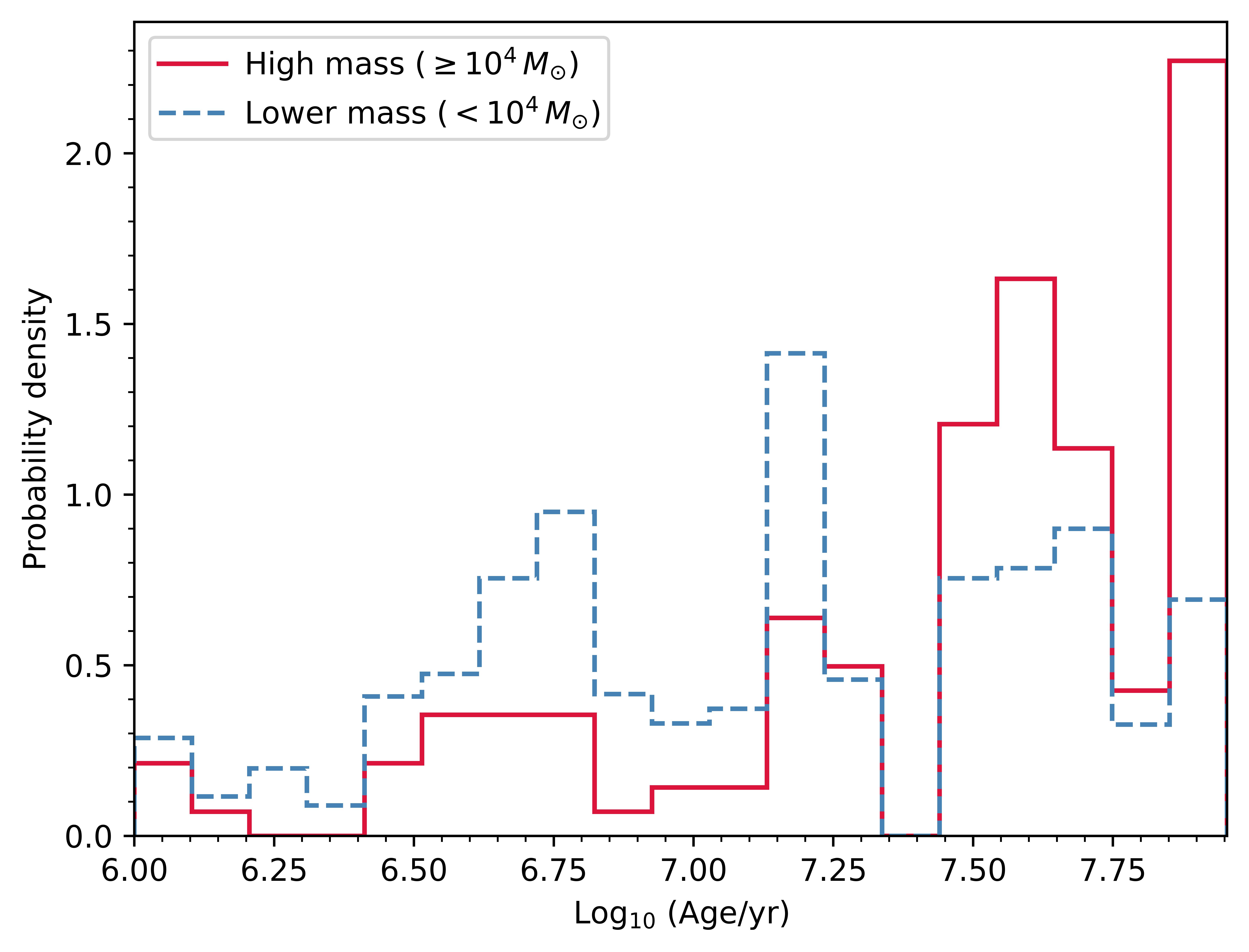}
\includegraphics[width=0.45\textwidth]{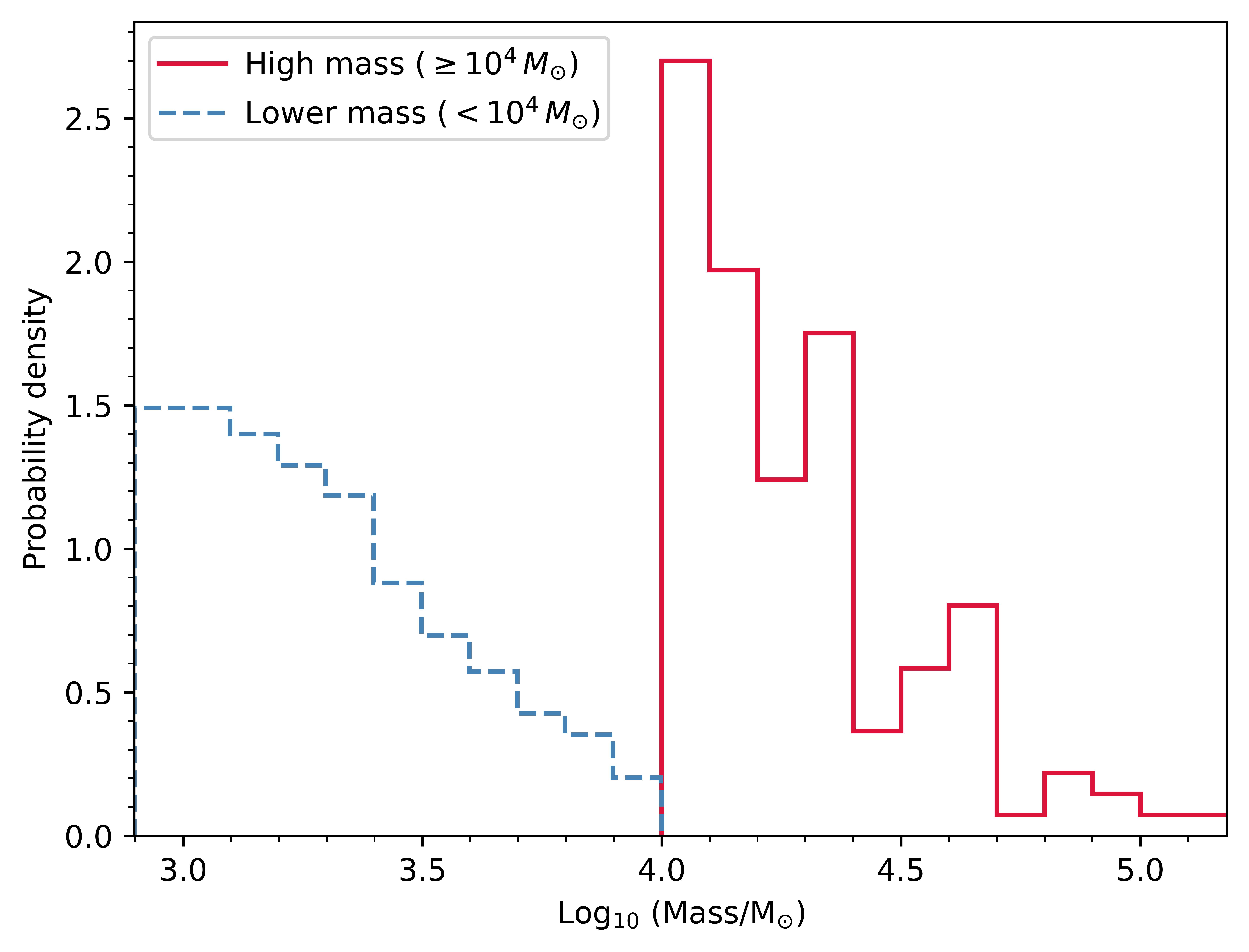}
\caption{Probability–density histograms for the mass–selected subsamples (“high” and “lower” mass): (left) $\log_{10}(\mathrm{age/yr})$ and (right) $\log_{10}(M/M_\odot)$. Both panels are normalised to unit area and use identical binning; the $y$–axis shows probability density. See Section~\ref{sect_results_mass}.\label{fig_hist_masses}}
\end{figure*}

We present diagnostic histograms of the age and mass distributions for the subsamples analysed in the main text. For each case, we show overlaid, probability–density histograms (unit area) of $\log_{10}(\mathrm{age/yr})$ and $\log_{10}(M/M_\odot)$ using identical binning to enable direct visual comparison. Specifically: (i) the eYSCs versus RECs comparison discussed in Section~\ref{sect_results_eYSCs} (Fig.~\ref{fig_hist_embedded}); (ii) the evolutionary subsamples (Young, Intermediate, Old) discussed in Section~\ref{sect_results_evo} (Fig.~\ref{fig_hist_ages}); and (iii) the mass–selected subsamples (“high” and “lower” mass) discussed in Section~\ref{sect_results_mass} (Fig.~\ref{fig_hist_masses}). Legends identify the subsamples, and the $y$–axis shows probability density in all panels. For completeness, Table~\ref{tab:medians_age_mass} summarises the median $\log_{10}(\mathrm{age/yr})$ and $\log_{10}(M/M_\odot)$ (with 16th–84th percentile ranges) and sample sizes for all subsamples used in this paper.

\begin{table}
\setlength{\tabcolsep}{4pt}        
\renewcommand{\arraystretch}{1.2}  
\centering
\caption{Median $\log_{10}(\mathrm{age/yr})$ and $\log_{10}(M/M_\odot)$ for each subsample, with 16th–84th percentile ranges. The final column lists the number of clusters used per subsample (identical for the age and mass estimates).}
\label{tab:medians_age_mass}
\begin{tabular}{lccc}
\hline
Sample & Age $\log_{10}(\mathrm{yr})$ & Mass $\log_{10}(M/M_\odot)$ & No. clusters \\
\hline
eYSC-NIR & $6.69^{+0.11}_{-0.52}$ & $2.96^{+0.42}_{-0.35}$ & 1084 \\
eYSC-Optical & $6.60^{+0.18}_{-0.55}$ & $2.96^{+0.38}_{-0.36}$ & 132\\
YSC & $7.48^{+0.82}_{-0.70}$ & $3.29^{+0.65}_{-0.65}$ & 5674\\
\hline
eYSCs & $6.69^{+0.15}_{-0.60}$ & $2.95^{+0.38}_{-0.36}$ & 750 \\
RECs  & $6.78^{+0.07}_{-0.18}$ & $2.71^{+0.49}_{-0.28}$ & 785 \\
\hline
Young        & $6.85^{+0.33}_{-0.24}$ & $2.83^{+0.51}_{-0.36}$ & 2805 \\
Intermediate & $7.70^{+0.20}_{-0.22}$ & $3.34^{+0.45}_{-0.30}$ & 1201 \\
Old & $8.48^{+0.22}_{-0.18}$  & $3.90^{+0.47}_{-0.34}$ & 1668 \\
\hline
High mass ($\geq 10^{4}\,M_{\odot}$) & $7.60^{+0.30}_{-0.77}$ & $4.24^{+0.31}_{-0.19}$ & 137 \\
Lower mass ($< 10^{4}\,M_{\odot}$)  & $7.18^{+0.52}_{-0.57}$ & $3.25^{+0.34}_{-0.24}$ & 2950 \\
\hline
\end{tabular}
\end{table}

\bsp	
\label{lastpage}
\end{document}